\documentclass[12pt]{article}
\usepackage[utf8]{inputenc}
\usepackage[margin=1in]{geometry}

\usepackage{amsmath}
\usepackage{amssymb}
\usepackage{enumerate}
\usepackage{graphicx}
\usepackage{natbib}
\usepackage{chapterbib}

\usepackage{tikz}
\usetikzlibrary{shapes.geometric}

\usepackage{dsfont}
\usepackage{subcaption}
\usepackage{breqn}

\usepackage{algpseudocode}
\usepackage{algorithm}

\usepackage{titling}
\newcommand{\ind}[1]{\mathds{1}(#1)}%
\newcommand{\subjectto}{\mbox{subject to}}   
\DeclareMathOperator{\B}{B}

\algnewcommand\algorithmicinput{\textbf{Input:}}
\algnewcommand\INPUT{\item[\algorithmicinput]}

\algnewcommand\algorithmicoutput{\textbf{Output:}}
\algnewcommand\OUTPUT{\item[\algorithmicoutput]}

\usepackage{booktabs,colortbl}
\definecolor{Gray}{gray}{0.9}

\usepackage{setspace}
\usepackage{filecontents}

\newif\ifnotblind
\notblindtrue

\title{Scaling Bayesian Probabilistic Record Linkage with Post-Hoc Blocking: An Application to the California Great Registers}
\ifnotblind
\author{Brendan S. McVeigh, Bradley T. Spahn , and Jared S. Murray}
\else
\fi

\begin{document}
    \maketitle

\begin{abstract}
\noindent{
Probabilistic record linkage (PRL) is the process of determining which records in two databases correspond to the same underlying entity in the absence of a unique identifier. Bayesian solutions to this problem provide a powerful mechanism for propagating uncertainty due to indeterminate links between records (via the posterior distribution over the linkage structures).  However, computational considerations severely limit the practical applicability of many existing Bayesian methods.  We propose a new computational approach that dramatically improves scalability of posterior inference, enabling Bayesian inference in problems orders of magnitude larger than many state-of-the-art algorithms. We use our methods to link a subset of an OCR'd dataset, the California Great Registers, containing hundreds of thousands of voter registrations spanning the New Deal realignment. Despite lacking a high quality blocking key our approach allows a posterior distribution to be estimated on a single machine in a matter of hours, enabling us to draw new inferences about party affiliation and switching from individual-level data over a critical time period in American politics.
\footnote{ Software in the form of a Julia package implementing our proposed method is publicly available at
\ifnotblind
https://github.com/brendanstats/BayesianRecordLinkage.jl
\else
(blinded)
\fi}

}
\end{abstract}

\section{Introduction}\label{sec:introduction}

Probabilistic record linkage (PRL) is the task of merging two or more databases that have entities in common but no unique identifier. Usually PRL must be done based on incomplete information -- attributes may be incorrectly or inconsistently recorded, or may be missing altogether.  Early developments of PRL were motivated by applications merging vital records and linking files from surveys and censuses to estimate population totals \citep{newcombe1959automatic,fellegi1969theory,jaro1989advances,copas1990record,winkler1991application}. Recent applications of PRL cover a wide range of problems, from linking health care data across providers \citep{dusetzina2014linking, sauleau2005medical, hof2017probabilistic} and following students across schools \citep{mackay2015educational, alicandro2017differences} to estimating casualty counts in conflicts \citep{price2014updated,chen2018,ball2019using}.

Bayesian approaches to PRL provide an appealing framework for uncertainty quantification and propagation via posterior sampling of the unknown links between records.  Bayesian methods have been deployed in a similar range of applied problems, including: Capture-recapture or multiple systems estimation, where the quantity of interest is a total population size \citep{liseo2011bayesian, tancredi2011hierarchical, steorts2016bayesian, sadinle2014detecting, sadinle2017bayesian, sadinle2018bayesian,betancourt2016, tancredi2018}, linking healthcare databases to estimate costs \citep{gutman2013bayesian}, and merging educational databases to study student outcomes \citep{dalzell2018regression}.  In this paper we use Bayesian PRL to link extracts of the California Great Registers -- historical voter registration rolls from the early 20th century \citep{spahn2018before} -- with the goal of understanding how patterns of party affiliation were changing at the individual-level in the New Deal era.

One drawback of Bayesian approaches to PRL is their computational burden. In the best of circumstances Bayesian inference can be computationally demanding, and making inference over a large discrete structure (the unobserved links between records) is particularly difficult.  Computational considerations have largely limited Bayesian inference for PRL to small problems, or large problems that can be effectively made small using clean quasi-identifiers through pre-processing steps known as indexing or blocking.  For example, we may only consider sets of records as potential matches if they agree on one or more quasi-identifiers such as a geographic area or a partial name match.

However, these pre-processing steps can lead to increased false non-match rates if the quasi-identifiers are subject to error. Moreover, many datasets lack the requisite clean quasi-identifying variables to make aggressive indexing or blocking feasible.  Our application in this paper is one such example: Like many historical datasets, the Great Registers contain few recorded attributes available to perform record linkage, all of which are subject to nontrivial amounts of error. It is also large; the files we work with contain over 260,000 records apiece.

To apply Bayesian modeling to this problem we introduce an approach to approximate Bayesian inference for PRL that is model agnostic and can provide samples from posterior distributions over links between datasets with hundreds of thousands of records in a matter of hours on a desktop computer. We realize these gains using a data-driven approach we call ``post-hoc blocking'' that limits attention to distinct sets of record pairs where there is significant ambiguity about true match status, and therefore the most to gain from posterior sampling.  Post-hoc blocking requires no labeled links/non-links (although it can utilize this data if available). Our methods build on previous work by 
\ifnotblind
\cite{mcveigh2017practical}
\else
(blinded citation)
\fi.

The paper proceeds as follows:  Section~\ref{sec:gr-intro} introduces the applied problem, linking voter registration files from the California Great Registers. Section~\ref{sec:background} collects background material and briefly reviews model-based approaches to record linkage, including the model we use for linking the Great Registers.  Section~\ref{sec:posthoc} introduces post-hoc blocking and describes its implementation. Sections~\ref{sec:alameda} presentes the results of linking the Great Register extract using our model and computational methods, comparing it to a recent proposal in the political science literature \citep{enamorado2018using}. Section~\ref{sec:switch_rates} anylazes patterns in party switching using our linked dataset. Section~\ref{sec:discussion} concludes with discussion and directions for future work.


\section{Linking the California Great Registers}\label{sec:gr-intro}

Beginning in 1900, California counties were required to publish a typeset copy of their voter lists in each election year \citep{spahn2018before}, known as the California Great Registers, which contain the name, address, party registration and occupation of every registered voter. This served as a record of the county's voters and as poll books on election day. The Great Registers provide a fine-grained tool for measuring the dynamics of partisan change over an especially interesting period of American history -- the New Deal realignment. From 1928 to 1936, a substantial number of Americans switched their partisan allegiance from the Republicans (the party of Herbert Hoover) to the Democrats (the party of Franklin Roosevelt). While this change is known to have taken place at the macro-level, the Great Registers are the first dataset that follows this change at the individual level, provided we can link individual voters over time.

One quantity of interest to historians and political scientists is the frequency with which voters changed party affiliation from 1932 to 1936, during Roosevelt's first term as president \citep{erikson1981partisan, andersen1979creation}. Decades of surveys conducted since 1948 have shown that  voters rarely switch parties. But this earlier period, before modern polling, featured the most dramatic and rapid change in partisanship in the twentieth century, making individual-level panel data from this period especially interesting.  In particular, individual-level panel data would enable a more detailed study of party switching behavior by demographic groups \citep{sundquist1983dynamics, corder2016counting, norpoth2019american}. To construct such a panel, we link records from the across the 1932 and 1936 registers based on the recorded  name, address and occupation.\footnote{While party might be an informative field in making a match, we withheld it from the matching process so that our estimate of the key quantity of interest -- the party switching rate -- is not biased toward stability.}

Though the structure of the data is relatively simple, transferring it from the printed page into digital format is challenging. Ancestry.com scanned and performed optical character recognition (OCR) on the Great Registers,  enabling use of the data by their subscribers for genealogical research. Since the quality of the scan as well as the original organization of the page can make the OCR fail to produce recognizable text or mistranscribe certain words and letters, much of the data was digitized imperfectly.  Once digitized, the data require further processing and standardization which is also subject to error.  These errors, coupled with natural variation (such as address changes and inconsistent recording of name variants and occupations), make linking challenging.

Because erroneous matches will inflate the match rate (a randomly selected voter from 1932 will share a party affiliation with a randomly selected voter from 1936 49\% of the time), making quality matches -- and accounting for uncertainty in which records match -- is essential to robustly estimating the party-switching rate. In addition to linking these two files, we also estimated false match rates to understand the performance of linkage algorithms and adjust our estimates, and compared the distribution of variables in the linked dataset to  cross-sections of the Great Registers to help gauge whether differential false non-matching might threaten the representativeness of our linked sample.    

\section{Model-Based Probabilistic Record Linkage}\label{sec:background}

We consider models for linking two files without any labeled true matching/non-matching pairs (``unsupervised'' models). In particular we assume two files which have no duplicates, a special but important use case of PRL. Here a record from one file can match at most one record from the other. The (at most) ``one-to-one'' matching assumption is often useful even when the input files are not perfectly deduplicated, as it can provide useful regularization of posterior match probability estimates.

To fix notation, suppose we have two sets of records, denoted $A$ and $B$, containing $n_A$ and $n_B$ records respectively.  
Records $a\in A$ and $b\in B$ are said to be ``matched'' or ``linked'' if they refer to the same underlying entity.  In this case the latent true match status for each record pair can be represented by an $n_A\times n_B$ binary matrix $C$, where
\begin{equation}
	C_{ab} = \left\{
\begin{array}{ll}
      1 & \text{if record $a$ matches record $b$} \\
      0 & \text{otherwise} \\
\end{array} 
\right.\label{eq:def_C}
\end{equation}
Since there are no duplicated records in $A$ or $B$, the matching between the two files must be one-to-one and the row and column sums of $C$ must all be less than or equal to one (i.e., not every record in each file need have a match). Our goal is to infer $C$ using observable attributes of the records.


\subsection{Comparing Records for PRL}\label{sec:background_comparisons}
For each record in the two files we obtain a set of attributes that partially identify the individual to whom the record corresponds. Common examples include names, addresses, or other demographic information.  In the Great Registers, these fields include name prefix, given name, surname, occupation, and address.   Pairs of records $(a,b)$ that appear similar on these attributes are more likely correspond to true matches ($C_{ab}=1$).  

\begin{table}
\centering
\captionsetup[subfigure]{justification=centering}
\begin{subfigure}{.32\textwidth}
\centering
\resizebox{0.98\textwidth}{!}{%
    \begin{tabular}{llllll}
    \hline
    \hline
    ID & First & MI & Last & Female & Occupation \\
    \hline
    $\mathrm{a}_1$ & gussie & \textemdash & albitz & 1 & \textemdash\\
    $\mathrm{a}_2$ & john & f & albitz & 0 & clerk\\
    $\mathrm{a}_3$ & hazel & m & dams & 1 & \textemdash \\
    $\mathrm{a}_4$ & herbert & a & dams & 0 & merchant \\
    \hline
    \hline
    \end{tabular}
}%
\caption{Data A}
\label{comparisons:dataa}
\end{subfigure}\hspace{.015\textwidth}%
\begin{subfigure}{.32\textwidth}
\centering
\resizebox{0.98\textwidth}{!}{%
    \begin{tabular}{llllll}
    \hline
    \hline
    ID & First & MI & Last & Female & Occupation \\
    \hline
    $\mathrm{b}_1$ & gussic & m & albitz & 1 & \textemdash\\
    $\mathrm{b}_2$ & john & f & albitz &  0 & railroad clerk\\
    $\mathrm{b}_3$ & hazel & m & adams & 1 & \textemdash\\
    $\mathrm{b}_4$ & herbert & a & adams & 0 & merchant \\
    \hline
    \hline
    \end{tabular}
}%
\caption{Data B}
\label{comparisons:datab}
\end{subfigure}\hspace{.015\textwidth}%
\begin{subfigure}{.32\textwidth}
\centering
\resizebox{0.98\textwidth}{!}{%
    \begin{tabular}{cccccc}
    \hline
    \hline
    Record Pair & First & MI & Last & Female & Occupation \\
    \hline
    $(\mathrm{a}_4, \mathrm{b}_1)$ & 0.00 & 0 & 0.47 & 0 & \textemdash \\
    $(\mathrm{a}_4, \mathrm{b}_2)$ & 0.46 & 0 & 0.47 & 1 & 0.42\\
    $(\mathrm{a}_4, \mathrm{b}_3)$ & 0.61 & 0 & 0.85 & 0 & \textemdash\\
    $(\mathrm{a}_4, \mathrm{b}_4)$ & 1.00 & 1 & 0.85 & 1 & 1.00\\
    \hline
    \hline
    \end{tabular}
}%
\caption{Record Comparisons}
\label{comparisons:examples}
\end{subfigure}
\caption{In (\subref{comparisons:dataa}) and (\subref{comparisons:datab}) are excerpts of two files $A$ and $B$. Table (\subref{comparisons:examples}) shows similarities between the last record in (\subref{comparisons:dataa}) and each record in (\subref{comparisons:datab}). String similarities are computed using a Jaro-Winkler string similarity with p = 0.1.} 
\label{tab:comparisons}
\end{table}

This notion is made explicit by directly modeling comparisons of these attributes between record pairs based on these attributes. The comparisons can be tailored to the specific features available. Table~\ref{tab:comparisons} provides a simple example of mapping record pairs to comparison vectors for a subset of the attributes available in the Great Registers.  The first two panels (\subref{comparisons:dataa} and \subref{comparisons:datab}) are records from file $A$ and file $B$, respectively.  Panel (\subref{comparisons:examples}) shows the constructed comparisons between the first entry in (\subref{comparisons:dataa}) and each entry in (\subref{comparisons:datab}). Here we compared surnames using a Jaro-Winkler similarity score \citep{winkler1990string}, ages using the absolute difference between the values, and counties using a strict matching criterion (1 if identical and 0 otherwise).  Many specialized comparison metrics have been developed. See \cite[Chapter 5]{christen2012data} 
for a detailed account of generating comparisons for different data types, and Section~\ref{sec:alameda_bayesian_comparison} for how we generated comparisons in the Great Registers data.

Suppose $d$ separate comparisons are generated for each record pair. We collect these in a vector:
\begin{equation}
    \gamma_{ab} = (\gamma_{ab}^1, \gamma_{ab}^2, \dots, \gamma_{ab}^d).\label{eq:def_gamma}
\end{equation}
and denote the collection of comparison vectors for all record pairs as $\Gamma$. These comparison vectors constitute the observed data in our model. Throughout we will assume that each $\gamma_{ab}^j$ is a {\em categorical} measure of similarity or agreement -- i.e., we follow others in the literature in discretizing continuous similarity measures (e.g. \citet{winkler1990string, larsen2010record, sadinle2017bayesian}) -- with higher levels corresponding to greater similarity.

\subsection{Model and Prior Specification} \label{sec:background_modelprior}

In this paper we use a Bayesian model and prior distribution introduced by \citet{fortini2001bayesian, fortini2002modelling}, generalized by \citet{larsen2005advances, larsen2010record} and studied further by \citet{sadinle2017bayesian} (although our computational approach is model agnostic). We use independent priors for linkage structure $C$ and as yet undefined model parameters.

We let $\pi(C)$ be a beta distribution for bipartite matchings \citep{larsen2005advances,larsen2010record,sadinle2017bayesian}, which we refer to as a ``Beta-bipartite'' distribution. This prior places a Beta-binomial distribution on the number of links $L = \sum_a\sum_b C_{ab}$ and -- given $L$ -- assigns a uniform prior over matching matrices $C$ with exactly $L$ links, so that
\[
\pi(C)  = \pi(C\mid L)\pi(L) = \frac{(n_B-L)!}{n_B!}\frac{\B(L+\alpha, n_A - L + \beta)}{\B(\alpha, \beta)}
\]
(assuming $n_A\leq n_B$, without loss of generality).

We adopt the standard mixture-model framework for modeling comparison vectors $\gamma_{ab}$, where the mixture components correspond to truly matching pairs  ($C_{ab}=1$) and non-matching pairs. (This formulation originates with \cite{fellegi1969theory}, discussed in the following subsection.) Define
\begin{equation}
m(g) = \Pr\left(\gamma_{ab} = g\mid C_{ab} = 1\right);\quad
u(g) = \Pr\left(\gamma_{ab} = g\mid C_{ab} = 0\right),\label{eq:def_muprob}
\end{equation}
for $g$ ranging over the possible values of the comparison vector. These parameters are often referred to as ``$m-$probabilities'' and ``$u-$probabilities'' in the literature, a convention we adopt here. We expect $m(g)>>u(g)$ when the comparison vector $g$ indicates significant agreement, and the reverse when $g$ indicates significant disagreement. 

The saturated model Eq.~\eqref{eq:def_muprob} is typically not estimable without restrictions on the parameter space. A common choice is conditional independence between comparisons. Define
\begin{align}
m_{jh} = \Pr\left(\gamma_{ab}^j = h|C_{ab} = 1\right);\quad
u_{jh} = \Pr\left(\gamma_{ab}^j = h|C_{ab} = 0\right),
\label{eq:def_mubinned}
\end{align}
for $1\leq j\leq d$ and $1\leq h\leq k_j$, where comparison $j$ has $k_j$ possible levels, and let $m_j=(m_{j1},\dots m_{jk_{j}})$ (with $u_j$ defined similarly).

Under conditional independence we have
\begin{equation}
		m(g) = \prod_{j=1}^d \prod_{h=1}^{k_j} m_{jh}^{\ind{g_j=h}};\quad 
		u(g) =  \prod_{j=1}^d \prod_{h=1}^{k_j} u_{jh}^{\ind{g_j=h}}. \label{eq:condindep}
\end{equation}
Less restrictive models impose log-linear or other constraints on the $m-$ and $u-$ probabilities \citep{thibaudeau1993discrimination,winkler1993improved,larsen2001iterative}.

Assuming prior independence between then $m-$ and $u-$ probabilities, we have the joint prior:
\begin{equation}
\pi(C, m_1,\hdots, m_d, u_1,\hdots, u_d)=\left(\prod_{A \times B} m(\gamma_{ab})^{C_{ab}}u(\gamma_{ab})^{1 - C_{ab}}\right)\left(\prod_{j=1}^d\pi(m_j)\pi(u_j)\right)\pi(C).\label{eq:model_big}
\end{equation}
Finally, we assume $m_j\sim \text{Dirichlet}(\alpha_{mj})$ and $u_j\sim \text{Dirichlet}(\alpha_{uj})$ where $\alpha_{mj}$ and $\alpha_{uj}$ are vectors of length $k_j$.

\subsection{Related work}
\subsubsection{The Fellegi-Sunter Framework}\label{sec:background_fellegi-sunter:introduction}

\cite{fellegi1969theory} provide an early approach to PRL using comparison vectors. The probability model they use is quite similar to the one presented above, but it assumes that the matching status of each record pair is independent of all the other pairs. In particular, \citet{fellegi1969theory} use the following two component mixture model:
\begin{equation}
\begin{gathered}
\Pr(C_{ab} = 1) = \pi \\
\Pr(\gamma_{ab} = g\mid C_{ab} = 1) = m(g),\quad \Pr(\gamma_{ab} = g\mid C_{ab} = 0) = u(g)\label{eq:fsmix}
\end{gathered}
\end{equation}
Conditional independence assumptions as in Eq~\eqref{eq:condindep} are commonly used in practice, in which case the model in Eq~\eqref{eq:fsmix} reduces to a two-component latent class model of the pairwise comparison vectors, with each comparison vector treated as an independent ``observation''.

\citet{fellegi1969theory} proposed estimating the model parameters via the method of moments, but it has become standard to use EM to maximize the likelihood function
\begin{equation}
L(m,u, \pi; \Gamma) = \prod_{(a,b)\in A\times B}
\pi m(\gamma_{ab}) + (1-\pi) u(\gamma_{ab})\label{eq:maxlik},
\end{equation}

\citep{winkler1988using}. Given values of the $m-$ and $u-$probabilities (or estimates thereof) \cite{fellegi1969theory} provide a decision rule for deriving a point estimate of $C$: Define a weight for each record pair:
\begin{equation}\label{eq:def_w}
w_{ab} = \log\left(\frac{m(\gamma_{ab})}{u(\gamma_{ab})}\right).
\end{equation}
To derive a point estimate of the linkage structure, any record pair $(a,b)$ with estimated weight $\hat w_{ab}>T_\mu$ is called a match ($\hat{C}_{ab} = 1$), and a record pair with $\hat w_{ab}<T_\lambda$ is called a non-match ($\hat C_{ab} =0$). If $T_\mu\neq T_\lambda$ then any remaining pairs have ``indeterminate'' match status and are evaluated manually. \cite{fellegi1969theory} describe how to set the thresholds $T_\mu$ and $T_\lambda$ to simultaneously control the false positive rate $\mu$ (the probability a non-matching pair is called a match) and false negative rate $\lambda$ (the probably a matching pair is called a non-match). \cite{fellegi1969theory} prove that their procedure minimizes the size of the indeterminate set for given values of $\mu$ and $\lambda$. 

As originally constructed, neither the methods for inferring $m$ and $u$ nor the decision rule for generating an estimate of the matching structure $C$ respect one-to-one matching constraints. \cite{jaro1989advances} proposed a three-stage approach for adapting the Fellegi-Sunter decision rule to respect one-to-one matching. The first stage generates estimates of $\hat m$ and $\hat u$ by maximizing \eqref{eq:maxlik}. The second stage generates $C^*$, an estimate of $C$ that satisfies the following assignment problem: 
 \begin{equation} \label{eq:jaroLSAP}
 \begin{aligned}
 C^* = \max_{C}   &\sum_{a,b\in A\times B} C_{ab}\hat w_{ab}& \\
 \subjectto\quad  & C_{ab} \in\{0, 1\}\\
 &\sum_{b\in B} \hspace{0.5em} C_{ab} = 1 \hspace{2em} \forall a\in A \\
 &\sum_{a\in A} \hspace{0.5em} C_{ab} \leq 1 \hspace{2em} \forall b\in B,
 \end{aligned}
 \end{equation}
where we assume that $n_A\leq n_B$.
Despite its combinatorial nature this optimization problem -- a linear sum assignment problem \citep{burkard2012assignment} -- can be solved relatively efficiently with standard linear programming techniques  (solving assignment problems is discussed in Section~\ref{sec:assignment_maxweights_thresholding} of the supplemental material). In the final stage, the point estimate  $\hat C$ is obtained from $C^*$ by setting $\hat C_{ab} = C_{ab}^*\ind{\hat w_{ab}\geq\lambda}$, where $\lambda$ plays the same role as in the FS decision rule.  (The algorithm presented in \cite{jaro1989advances} has an error, which we describe how to correct in supplemental material (Section~\ref{sec:assignment}).)

\subsubsection{Bayesian Modeling for Probabilistic Record Linkage}\label{sec:background_bayesPRLl}

Many approaches to Bayesian PRL utilize the same comparison-vector based model we present in Section~\ref{sec:background_modelprior} (e.g. \citep{fortini2001bayesian, fortini2002modelling, larsen2005advances, larsen2010record, sadinle2017bayesian}). \cite{sadinle2017bayesian} provides a loss function that leads to a Bayes estimate that is similar to the Fellegi-Sunter decision rule. Other Bayesian models avoid the reduction to comparison vectors by modeling population distributions of fields and error-generating processes directly \citep{tancredi2011hierarchical,tancredi2013accounting,steorts2015entity,steorts2016bayesian,marchant2019d} and/or specify joint models for $C$ and the ultimate analysis of interest, such as a regression model where the response variable is only available on one of the two files \citep{gutman2013bayesian,dalzell2017file}.

Most implementations of Bayesian PRL under one-to-one matching update $C$ using local Metropolis-Hastings moves.  At each step the algorithm proposes to either add or drop individual links, or swap the links between two record pairs, as described in e.g. \cite{fortini2002modelling,larsen2005advances,green2006bayesian,tancredi2011hierarchical}.\footnote{More efficient samplers have been defined for specific models and priors not subject to the one-to-one constraint, e.g. \citep{betancourt2016, marchant2019d} }  These algorithms do not lend themselves to large datasets, as they inefficiently explore the posterior distribution when the number of potential links is large.  A notable exception is \cite{zanella2019informed}, in which the current values of model parameters are used to make more efficient local MCMC proposals.  However, the computational cost of these informed moves is nontrivial (as discussed in Section~\ref{sec:posthoc}).   

MCMC algorithms can be more effective when the records in both files can be partitioned or ``blocked'' such that links between records can only occur within elements of the partition (the blocks). Indeed, with high-quality blocking variables some of these blocks can be small enough to enumerate, which admits simpler Gibbs sampling updates of the corresponding submatrices of $C$  \citep{gutman2013bayesian,dalzell2018regression}. \cite{gutman2013bayesian} are able to scale to millions of records with the availability of several such blocking variables. 

However, often high-quality blocking variables are unavailable. In these cases other authors have incorporated ideas from blocking in deriving proposal distributions or defining priors. \cite{zanella2019informed} incorporates random blocking into their proposal distribution for large datasets (which we discuss further below). \cite{marchant2019d} introduce a partition of latent true entities which define their prior over linkage structures, which is similar in spirit to a latent blocking scheme (although in its most general form it does not strictly limit which observed records can be linked to other observed records).  In combination with distributed computing, model-specific collapsed Gibbs sampling, and some approximations tailored to their particular model and prior, this yields a more scalable approach to Bayesian PRL.

In contrast, our ``post-hoc'' blocking method (outlined in the next subsection) uses a generic data-driven approach to derive compact blocks in the absence of high-quality blocking variables.  It does not depend on the particular prior distribution or model under consideration and delivers computationally efficient inference even without distributed architectures, parallel computing, or additional MCMC sampling tricks or approximations (although it can benefit from these when available).

\section{Post-Hoc Blocking For Bayesian Record Linkage}\label{sec:posthoc}

Probabilistic record linkage is inherently computationally expensive; with files of size $n_A$ and $n_B$ there are $n_An_B$ record comparisons to be made.  It is generally necessary to reduce the number of record pairs under consideration during data pre-processing, a process known as {\em indexing} or {\em blocking}. We provide a short overview of traditional pre-processing steps before introducing our new strategy of ``post-hoc blocking'' for scaling Bayesian PRL.

\subsection{Traditional Approaches: Indexing, Blocking and Filtering}\label{sec:posthoc_trad-blocking}

Traditional approaches to reducing the number of potentially-matching record pairs can be separated into three categories: indexing, blocking, and filtering \citep{murray2016probabilistic}.  Indexing refers to any technique for excluding a record pair from consideration before performing a complete comparison; for example, we might exclude any record pairs with years of birth that differ by more than five. A blocking scheme is an indexing scheme that requires candidate record pairs to agree on a single derived comparison, known as a {\em blocking key}. This induces a nontrivial partition of the records such that all links must occur within elements of the partition (the blocks). For example, indexing by requiring that matching record pairs agree on a county and first initial defines a blocking scheme, since matches can only occur within blocks defined by the cross-product of letters and counties.  Filtering refers to excluding record pairs {\em after} a complete comparison has been made. Little reference is made to filtering in the literature, but it is featured in many implementations of PRL (e.g. the U.S. Census Bureau's BigMatch software \citep{yancey2002bigmatch}).

Reducing the number of comparisons by blocking is particularly attractive since it effectively leads to a collection of smaller PRL problems that can be solved in parallel. However, it is relatively rare to have a single blocking key that leads to an effective reduction in the number of candidate pairs without excluding many true matches in the process. Indexing is more common in practice  (see \citep{steorts2014comparison} and Chapter 5 of \cite{christen2012data} for recent reviews of indexing methods). A common indexing scheme is to combine of multiple blocking passes (see e.g. \cite{winkler2010fast} for a high-profile example), which \cite{murray2016probabilistic} calls indexing by disjunctions (of blocking keys). For example, we might include all record pairs that come from the same county {\em or} match on the first three characters of their first name. Indexing by disjunctions does not itself yield a blocking scheme, however. In the next subsection we describe a data-driven procedure for efficiently constructing blocks {\em after} indexing, without requiring a set of labeled matching/non-matching record pairs.

\subsection{Post-Hoc Blocking for Bayesian PRL} \label{sec:posthoc_bayesPRL}

Given their computational complexity, Bayesian implementations of PRL can benefit from stricter blocking or indexing than other methods.  Stricter indexing increases the risk of false non-matches, so it is important that the indexing be as efficient as possible by admitting plausibly matching record pairs while excluding clearly non-matching pairs.  We propose constructing a high-quality blocking key from the available fields which separates most of the obviously non-matching pairs across blocks while keeping plausible matches within the same block. This is difficult to do via traditional indexing {\em a priori}, especially in the absence of labeled matches, so we suggest a data-driven approaches to choose the blocks (with or without labeled matching and non-matching pairs) -- hence the name  ``post-hoc blocking''. 

Post-hoc blocking is straightforward: First, perform traditional blocking or indexing only to the extent necessary to make computing comparison vectors feasible. Second, estimate matching weights or probabilities for each record pair.  We refer to these generically as post-hoc blocking weights.  The only criteria for these weights is that they reliably give relatively high weight to plausible matching pairs and low weight to true non-matching pairs; they need not be well-calibrated probabilities or proper likelihood ratios. Third, conduct an additional blocking pass using the estimated weights to construct the blocking key, reducing the number of record pairs just enough to make running an MCMC algorithm feasible.   With the post-hoc blocks in hand, we run an MCMC algorithm as usual, restricting the proposal distribution to only consider matches within post-hoc blocks. 

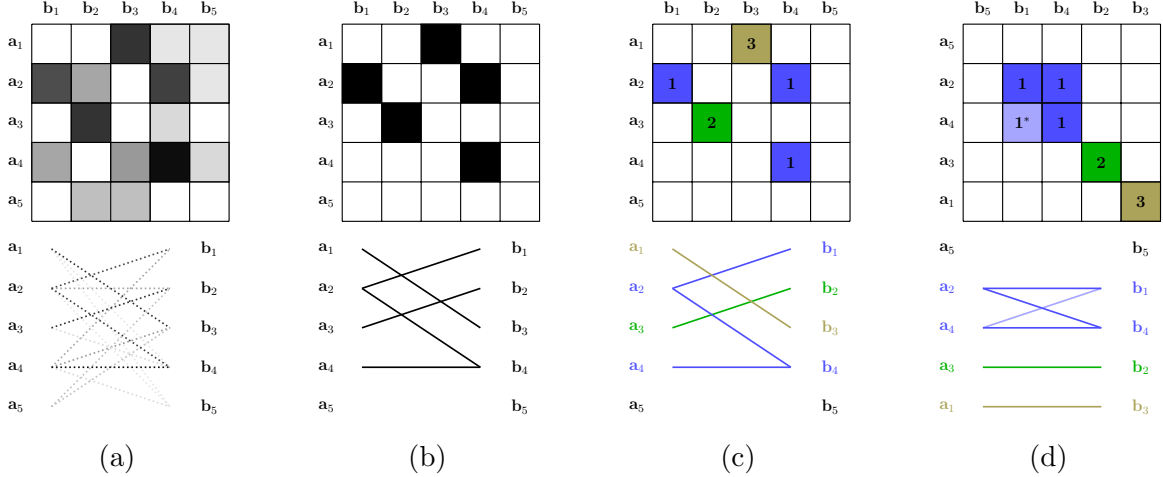
\begin{figure*}
\centering
\captionsetup[subfigure]{justification=centering}
\begin{subfigure}{.2\textwidth}
\centering
\resizebox{0.95\textwidth}{!}{%
\begin{tikzpicture}
\filldraw[fill=black!35!white] (0,1) rectangle (1,2);
\filldraw[fill=black!35!white] (1,3) rectangle (2,4);
\filldraw[fill=black!40!white] (2,1) rectangle (3,2);
\filldraw[fill=black!15!white] (3,2) rectangle (4,3);
\filldraw[fill=black!15!white] (4,1) rectangle (5,2);
\filldraw[fill=black!25!white] (1,0) rectangle (3,1);
\filldraw[fill=black!10!white] (3,3) rectangle (5,5);

\filldraw[fill=black!70!white] (0,3) rectangle (1,4);
\filldraw[fill=black!80!white] (1,2) rectangle (2,3);
\filldraw[fill=black!80!white] (2,4) rectangle (3,5);
\filldraw[fill=black!75!white] (3,3) rectangle (4,4);
\filldraw[fill=black!95!white] (3,1) rectangle (4,2);

\node[] at (-0.4,4.5) {$\mathbf{a}_1$};
\node[] at (-0.4,3.5) {$\mathbf{a}_2$};
\node[] at (-0.4,2.5) {$\mathbf{a}_3$};
\node[] at (-0.4,1.5) {$\mathbf{a}_4$};
\node[] at (-0.4,0.5) {$\mathbf{a}_5$};

\node[] at (0.5,5.4) {$\mathbf{b}_1$};
\node[] at (1.5,5.4) {$\mathbf{b}_2$};
\node[] at (2.5,5.4) {$\mathbf{b}_3$};
\node[] at (3.5,5.4) {$\mathbf{b}_4$};
\node[] at (4.5,5.4) {$\mathbf{b}_5$};

\draw[step=1cm,black,thick] (0,0) grid (5,5);

\node[circle] at (-0.4, -0.7) {$\mathbf{a}_1$};
\node[circle] at (-0.4, -1.7) {$\mathbf{a}_2$};
\node[circle] at (-0.4, -2.7) {$\mathbf{a}_3$};
\node[circle] at (-0.4, -3.7) {$\mathbf{a}_4$};
\node[circle] at (-0.4, -4.7) {$\mathbf{a}_5$};

\node[circle] at (4.5, -0.7) {$\mathbf{b}_1$};
\node[circle] at (4.5, -1.7) {$\mathbf{b}_2$};
\node[circle] at (4.5, -2.7) {$\mathbf{b}_3$};
\node[circle] at (4.5, -3.7) {$\mathbf{b}_4$};
\node[circle] at (4.5, -4.7) {$\mathbf{b}_5$};

\draw [very thick, dotted, black!35!white] (0.5, -3.7) -- (3.5, -0.7);
\draw [very thick, dotted, black!35!white] (0.5, -1.7) -- (3.5, -1.7);
\draw [very thick, dotted, black!40!white] (0.5, -3.7) -- (3.5, -2.7);
\draw [very thick, dotted, black!15!white] (0.5, -2.7) -- (3.5, -3.7);
\draw [very thick, dotted, black!15!white] (0.5, -3.7) -- (3.5, -4.7);
\draw [very thick, dotted, black!25!white] (0.5, -4.7) -- (3.5, -1.7);
\draw [very thick, dotted, black!25!white] (0.5, -4.7) -- (3.5, -2.7);
\draw [very thick, dotted, black!10!white] (0.5, -1.7) -- (3.5, -4.7);
\draw [very thick, dotted, black!10!white] (0.5, -0.7) -- (3.5, -3.7);
\draw [very thick, dotted, black!10!white] (0.5, -0.7) -- (3.5, -4.7);

\draw [very thick, dotted, black!70!white] (0.5, -1.7) -- (3.5, -0.7);
\draw [very thick, dotted, black!80!white] (0.5, -2.7) -- (3.5, -1.7);
\draw [very thick, dotted, black!80!white] (0.5, -0.7) -- (3.5, -2.7);
\draw [very thick, dotted, black!75!white] (0.5, -1.7) -- (3.5, -3.7);
\draw [very thick, dotted, black!95!white] (0.5, -3.7) -- (3.5, -3.7);

\end{tikzpicture}
}%
\caption{}
\label{cluster:weight}
\end{subfigure}\hspace{.05\linewidth}%
\begin{subfigure}{.2\textwidth}
\centering
\resizebox{0.95\textwidth}{!}{%
\begin{tikzpicture}
\filldraw[fill=black] (0,3) rectangle (1,4);
\filldraw[fill=black] (1,2) rectangle (2,3);
\filldraw[fill=black] (2,4) rectangle (3,5);
\filldraw[fill=black] (3,3) rectangle (4,4);
\filldraw[fill=black] (3,1) rectangle (4,2);

\node[] at (-0.4,4.5) {$\mathbf{a}_1$};
\node[] at (-0.4,3.5) {$\mathbf{a}_2$};
\node[] at (-0.4,2.5) {$\mathbf{a}_3$};
\node[] at (-0.4,1.5) {$\mathbf{a}_4$};
\node[] at (-0.4,0.5) {$\mathbf{a}_5$};

\node[] at (0.5,5.4) {$\mathbf{b}_1$};
\node[] at (1.5,5.4) {$\mathbf{b}_2$};
\node[] at (2.5,5.4) {$\mathbf{b}_3$};
\node[] at (3.5,5.4) {$\mathbf{b}_4$};
\node[] at (4.5,5.4) {$\mathbf{b}_5$};

\draw[step=1cm,black,thick] (0,0) grid (5,5);

\node[circle] at (-0.4, -0.7) {$\mathbf{a}_1$};
\node[circle] at (-0.4, -1.7) {$\mathbf{a}_2$};
\node[circle] at (-0.4, -2.7) {$\mathbf{a}_3$};
\node[circle] at (-0.4, -3.7) {$\mathbf{a}_4$};
\node[circle] at (-0.4, -4.7) {$\mathbf{a}_5$};

\node[circle] at (4.5, -0.7) {$\mathbf{b}_1$};
\node[circle] at (4.5, -1.7) {$\mathbf{b}_2$};
\node[circle] at (4.5, -2.7) {$\mathbf{b}_3$};
\node[circle] at (4.5, -3.7) {$\mathbf{b}_4$};
\node[circle] at (4.5, -4.7) {$\mathbf{b}_5$};

\draw [very thick, black] (0.5, -1.7) -- (3.5, -0.7);
\draw [very thick, black] (0.5, -2.7) -- (3.5, -1.7);
\draw [very thick, black] (0.5, -0.7) -- (3.5, -2.7);
\draw [very thick, black] (0.5, -1.7) -- (3.5, -3.7);
\draw [very thick, black] (0.5, -3.7) -- (3.5, -3.7);

\end{tikzpicture}
}%
\caption{}
\label{cluster:link}
\end{subfigure}\hspace{.05\linewidth}%
\begin{subfigure}{.2\textwidth}
\centering
\resizebox{0.95\textwidth}{!}{%
\begin{tikzpicture}
\filldraw[fill=blue!70!white] (0,3) rectangle (1,4);
\filldraw[fill=blue!70!white] (3,3) rectangle (4,4);
\filldraw[fill=blue!70!white] (3,1) rectangle (4,2);
\filldraw[fill=green!70!black] (1,2) rectangle (2,3);
\filldraw[fill=yellow!60!black] (2,4) rectangle (3,5);

\node[] at (0.5,3.5) {$\mathbf{1}$};
\node[] at (3.5,3.5) {$\mathbf{1}$};
\node[] at (3.5,1.5) {$\mathbf{1}$};
\node[] at (1.5,2.5) {$\mathbf{2}$};
\node[] at (2.5,4.5) {$\mathbf{3}$};

\node[] at (-0.4,4.5) {$\mathbf{a}_1$};
\node[] at (-0.4,3.5) {$\mathbf{a}_2$};
\node[] at (-0.4,2.5) {$\mathbf{a}_3$};
\node[] at (-0.4,1.5) {$\mathbf{a}_4$};
\node[] at (-0.4,0.5) {$\mathbf{a}_5$};

\node[] at (0.5,5.4) {$\mathbf{b}_1$};
\node[] at (1.5,5.4) {$\mathbf{b}_2$};
\node[] at (2.5,5.4) {$\mathbf{b}_3$};
\node[] at (3.5,5.4) {$\mathbf{b}_4$};
\node[] at (4.5,5.4) {$\mathbf{b}_5$};

\draw[step=1cm,black,thick] (0,0) grid (5,5);

\node[circle, yellow!60!black] at (-0.4, -0.7) {$\mathbf{a}_1$};
\node[circle, blue!70!white] at (-0.4, -1.7) {$\mathbf{a}_2$};
\node[circle, green!70!black] at (-0.4, -2.7) {$\mathbf{a}_3$};
\node[circle, blue!70!white] at (-0.4, -3.7) {$\mathbf{a}_4$};
\node[circle] at (-0.4, -4.7) {$\mathbf{a}_5$};

\node[circle, blue!70!white] at (4.5, -0.7) {$\mathbf{b}_1$};
\node[circle, green!70!black] at (4.5, -1.7) {$\mathbf{b}_2$};
\node[circle, yellow!60!black] at (4.5, -2.7) {$\mathbf{b}_3$};
\node[circle, blue!70!white] at (4.5, -3.7) {$\mathbf{b}_4$};
\node[circle] at (4.5, -4.7) {$\mathbf{b}_5$};

\draw [very thick, blue!70!white] (0.5, -1.7) -- (3.5, -0.7);
\draw [very thick, green!70!black] (0.5, -2.7) -- (3.5, -1.7);
\draw [very thick, yellow!60!black] (0.5, -0.7) -- (3.5, -2.7);
\draw [very thick, blue!70!white] (0.5, -1.7) -- (3.5, -3.7);
\draw [very thick, blue!70!white] (0.5, -3.7) -- (3.5, -3.7);

\end{tikzpicture}
}%
\caption{}
\label{cluster:label}
\end{subfigure}\hspace{.05\linewidth}%
\begin{subfigure}{.2\textwidth}
\centering
\resizebox{0.95\textwidth}{!}{%
\begin{tikzpicture}
\filldraw[fill=blue!70!white] (1,3) rectangle (2,4);
\filldraw[fill=blue!35!white] (1,2) rectangle (2,3);
\filldraw[fill=blue!70!white] (2,3) rectangle (3,4);
\filldraw[fill=blue!70!white] (2,2) rectangle (3,3);
\filldraw[fill=green!70!black] (3,1) rectangle (4,2);
\filldraw[fill=yellow!60!black] (4,0) rectangle (5,1);

\node[] at (2.5,3.5) {$\mathbf{1}$};
\node[] at (2.5,2.5) {$\mathbf{1}$};
\node[] at (1.5,3.5) {$\mathbf{1}$};
\node[] at (1.5,2.5) {$\mathbf{1^*}$};
\node[] at (3.5,1.5) {$\mathbf{2}$};
\node[] at (4.5,0.5) {$\mathbf{3}$};

\node[] at (-0.4,4.5) {$\mathbf{a}_5$};
\node[] at (-0.4,3.5) {$\mathbf{a}_2$};
\node[] at (-0.4,2.5) {$\mathbf{a}_4$};
\node[] at (-0.4,1.5) {$\mathbf{a}_3$};
\node[] at (-0.4,0.5) {$\mathbf{a}_1$};

\node[] at (0.5,5.4) {$\mathbf{b}_5$};
\node[] at (1.5,5.4) {$\mathbf{b}_1$};
\node[] at (2.5,5.4) {$\mathbf{b}_4$};
\node[] at (3.5,5.4) {$\mathbf{b}_2$};
\node[] at (4.5,5.4) {$\mathbf{b}_3$};

\draw[step=1cm,black,thick] (0,0) grid (5,5);

\node[circle] at (-0.4, -0.7) {$\mathbf{a}_5$};
\node[circle, blue!70!white] at (-0.4, -1.7) {$\mathbf{a}_2$};
\node[circle, blue!70!white] at (-0.4, -2.7) {$\mathbf{a}_4$};
\node[circle, green!70!black] at (-0.4, -3.7) {$\mathbf{a}_3$};
\node[circle, yellow!60!black] at (-0.4, -4.7) {$\mathbf{a}_1$};

\node[circle] at (4.5, -0.7) {$\mathbf{b}_5$};
\node[circle, blue!70!white] at (4.5, -1.7) {$\mathbf{b}_1$};
\node[circle, blue!70!white] at (4.5, -2.7) {$\mathbf{b}_4$};
\node[circle, green!70!black] at (4.5, -3.7) {$\mathbf{b}_2$};
\node[circle, yellow!60!black] at (4.5, -4.7) {$\mathbf{b}_3$};

\draw [very thick, blue!35!white] (0.5, -2.7) -- (3.5, -1.7);
\draw [very thick, blue!70!white] (0.5, -1.7) -- (3.5, -1.7);
\draw [very thick, blue!70!white] (0.5, -2.7) -- (3.5, -2.7);
\draw [very thick, blue!70!white] (0.5, -1.7) -- (3.5, -2.7);
\draw [very thick, green!70!black] (0.5, -3.7) -- (3.5, -3.7);
\draw [very thick, yellow!60!black] (0.5, -4.7) -- (3.5, -4.7);

\end{tikzpicture}
}%
\caption{}
\label{cluster:order}
\end{subfigure}  
\caption{An example of \textit{post-hoc blocking}, weight/adjacency matrices are shown in the top row with the corresponding bipartite graphs shown in the bottom row. (\subref{cluster:weight}) Shows an example of estimated post-hoc blocking weights (dark cells corresponding to larger weights). (\subref{cluster:link}) A binary matrix constructed by thresholding the weight matrix at $w_0$. Edges in the graph correspond to possible matches. (\subref{cluster:label}) We number and color the connected components of the graph; these will form the post-hoc blocks.  (\subref{cluster:order}) We reorder the records to group them into the post-hoc blocks. Note that record pair $(a_4,b_1)$ (labeled $1^*$) is included to complete post-hoc block one, even though its weight was below $w_0$.} 
\label{fig:cluster}
\end{figure*}

Figures~\ref{cluster:weight} - \ref{cluster:order} illustrate the process of post-hoc block generation.  The rows and columns of the heatmaps correspond to records from file $A$ and file $B$, respectively. Figure ~\ref{cluster:weight} shows a heatmap of the post-hoc blocking weights for each pair, with darker squares signifying larger weights. To generate a set of post-hoc blocks we begin by thresholding the matrix of weights at a low value $w_0$.  Figure~\ref{cluster:link} shows the thresholded matrix, where black boxes correspond to the record pairs with weights over the threshold. 

At this point we have defined a bipartite graph between the records in files $A$ and $B$ (shown below the heatmap); an edge is present between records $a_i$ and $b_j$ if the weight for the record pair exceeds $w_0$. The sets of records corresponding to the nodes in each connected component are the {\em post-hoc blocks}; these are labeled in Figures~\ref{cluster:label} and \ref{cluster:order}. Post-hoc blocking significantly reduces the number of candidate pairs  while identifying a block of records that appear to have multiple plausible configurations (post-hoc block 1, in blue). After a first pass, if any of the remaining post-hoc blocks are too large, we increase $w_0$ and apply this procedure recursively within the large blocks.

The procedure for sampling from an approximate posterior distribution for $C$ employing post-hoc blocking is summarized in Algorithm~\ref{alg:posthoc} below; implementation details follow.

\begin{algorithm}
\caption{Post-hoc Blocking with Restricted MCMC}
\label{alg:posthoc}
\begin{algorithmic}
\INPUT{Comparison vectors $\Gamma$ for a set of record pairs, initial weight threshold $w_{min}$, maximum post-hoc block size $N_c$}
\OUTPUT{Approximate posterior distribution for $C$ and other model parameters}
  \begin{enumerate}
  \item Estimate post-hoc blocking weights $\hat{w}_{ab}$.
  \item Compute the matrix $E$ where $e_{ab} = \ind{\hat{w}_{ab} > w_0}$ with $w_0 = w_{min}$
  \item Find the connected components of $G$, where $G$ is defined as the bipartite graph with adjacency matrix $E$. The set of records corresponding to the nodes in each connected component are the { post-hoc blocks}
  \item For post-hoc blocks larger than $N_c$ repeat 2. and 3. with a threshold $w_0' > w_0$.  Apply recursively on any resulting post-hoc blocks larger than $N_c$
  \item Run a standard MCMC algorithm, fixing $C_{ab}=0$ for all record pairs outside of the post-hoc blocks.
  \end{enumerate}
\end{algorithmic}
\end{algorithm}

\noindent
{\bf Weight estimation.} Clearly the performance of post-hoc blocking will depend on the quality of the weights.  However, compared to using the weights to {\em identify} truly matching pairs, for the purposes of post-hoc blocking we can tolerate lower quality weights.   What is essential is that they give high weight to truly matching and ambiguous record pairs, while giving low weight to clearly non-matching pairs (so the blocks are compact).  It is less important that the weights give a good rank ordering of the truly matching/ambiguous pairs -- as long as they end up in the same post-hoc block, the Bayesian model and MCMC algorithm will treat them appropriately.

If labeled true matching and non-matching record pairs are available we could use these to predict matching probabilities for the remaining pairs using standard classification methods. These predicted probabilities will often fail to be well calibrated; for example, when a record in file A has multiple plausible candidates in file B they may all receive high matching probabilities with a classifier trained treating record pairs as iid observations. Regardless, these records will be gathered into the same post-hoc block and the uncertainty in the matching structure will be accurately represented in the posterior distribution.

Alternatively, in the absence of labeled record pairs we could use EM estimates of the Fellegi-Sunter weights (Eq~\eqref{eq:def_w}) as post-hoc blocking weights. This can work well in settings where there are many attributes available for matching and where most of the records in $A$ also appear in $B$ \citep{winkler2002methods}.  When the two files have few fields in common, or there is significant error in the fields available, or when there is limited overlap between the files, EM-estimated weights can perform quite poorly \citep{winkler2002methods,tancredi2011hierarchical,sadinle2017bayesian}. More reliable weights can be obtained by getting coarse estimates of $m-$ and $u-$ probabilities while accounting for the one-to-one matching constraint.  We outline a method for obtaining such weights using a novel penalized likelihood procedure in Section~\ref{sec:assignment_maxweights} of the supplemental material; this is how we generate the post-hoc blocking weights for our application in Section~\ref{sec:alameda}.

\noindent
{\bf Obtaining post-hoc blocks.}  For a given threshold $w_0$, finding the post-hoc blocks is equivalent to finding the connected components of a bipartite graph. This is a well-studied problem with computationally efficient solutions \citep{tarjan1972depth,gazit1986optimal}. 

\noindent
{\bf Selecting the maximum block size $N_c$}. 
Choosing the maximum block size $N_c$ requires balancing statistical accuracy (the quality of our posterior approximation) against computational efficiency. Smaller values of $N_c$ are more likely to exclude true matching pairs, increasing the false non-match rate. Excluding truly non-matching pairs which are not {\em obviously} non-matches also risks misrepresenting posterior uncertainty. On the other hand, selecting a larger $N_c$ decreases bias by admitting more record pairs and yields a smaller number of post-hoc blocks of larger size. Larger post-hoc blocks lead to increased computation time, as the most significant computational gains accrue when a large fraction of the post-hoc blocks are small.  Given these considerations we should choose the largest $N_c$ that leads to a computationally feasible MCMC algorithm. What constitutes a ``feasible'' problem will naturally be context dependent.

\noindent
{\bf Implementing restricted MCMC algorithms.}  
Post-hoc blocking can achieve massive reductions in scale  relative to traditional blocking schemes.  Generally, a large number of small or singleton blocks are produced, in addition to a smaller number of larger blocks. This distribution of block sizes makes possible a restricted MCMC algorithm which mixes much more efficiently than standard approaches.

For very small blocks we perform Gibbs updates by enumerating all possible values of the corresponding submatrix of $C$ and sampling proportional to their unnormalized posteriors.  Other implementations of Bayesian PRL have taken advantage of this enumerability when a large number of high-quality traditional blocking fields are available (e.g.  \citet{gutman2013bayesian}). However, post-hoc blocking is more likely to produce a large number of small blocks than traditional blocking, especially in the absence of one or more high-quality blocking keys.

For moderately-sized blocks, informative locally balanced Metropolis-Hastings proposals can be used instead of simple add/drop/swap proposals \citep{zanella2019informed}. \cite{zanella2019informed} showed that locally balanced proposals can dramatically improve mixing over standard Metropolis-Hastings proposals in Bayesian PRL models. However, locally balanced proposals also become prohibitively costly for large blocks: For a $k_A\times k_B$ block containing $L$ links at one iteration, the likelihood (up to a constant) must be computed $2(k_Ak_B - L(L- 1))$ times to perform a single locally balanced update.  \cite{zanella2019informed} mitigated this issue by including random sub-block generation as part of the locally balanced proposal. But as the file sizes increase these completely random sub-blocks are increasingly unlikely to capture all or even many of the plausible candidates for each record in the block, increasing mixing time. In contrast, our post-hoc blocks are specifically constructed to capture all the plausible candidates for a given record in the same compact block. 

The integration of post-hoc blocking with locally balanced moves and Gibbs updates produces an MCMC algorithm which mixes substantially faster for large problems than standard approaches.  However, the posterior distribution obtained under post-hoc blocking is only an approximation, as the posterior probability of links between record pairs outside of the post-hoc blocks is artificially set to zero.\footnote{At the cost of significant additional bookkeeping it is possible in principle to include the post-hoc blocking thresholds as part of a proposal distribution similar to \cite{zanella2019informed}; we leave this extension for future work.}

In small problems where we can check against the full posterior the practical effect of this approximation seems to be limited (see Section~\ref{sec:assignment_italiam-census} in the supplemental material for an example).  In large problems, an approximation of some sort seems unavoidable -- it is infeasible to run any MCMC algorithm over datasets with hundreds of thousands of records generating hundreds of millions or billions of candidate record pairs (after indexing) sufficiently long to mix properly. The result is that in a practical MCMC run the vast majority of those entries we fix at zero would have posterior probabilities estimated at or near zero anyway. With post-hoc blocking and restricted MCMC using locally balanced proposals we are able to hone in on areas of non-negligible posterior uncertainty, and spend more of our time sampling in these regions.

\subsection{Post-hoc Blocking Versus Traditional Blocking/Indexing/Filtering}\label{sec:posthoc_phb-trad}

Post-hoc blocking combines ideas from indexing (specifically blocking) and filtering. However, it is not a special case of either. In traditional indexing and blocking, the goal is to avoid a complete comparison of the record pairs. As a result, the record pairs excluded by indexing are simply ignored and have no impact on model fitting. The same is typically true under filtering -- the record pairs that are filtered {\em after} a complete comparison have been made are ignored during model fitting, even though their comparison vectors are available. 

In post-hoc blocking we use of all the generated comparison vectors by fixing $C_{ab}=0$ for record pairs outside the post-hoc blocks. Even though they cannot be matched, data from these record pairs are used to estimate model parameters. We take a similar approach to record pairs excluded by the initial blocking/indexing scheme -- although their comparison vectors are not available, we can compute the relevant summary statistics exactly under a conditional independence model (see Appendix~\ref{sec:ucorrection} for details). This avoids some of the more pernicious bias-inducing effects of blocking, indexing, and filtering on subsequent parameter estimation described by \cite{murray2016probabilistic}.  In the model introduced in Section~\ref{sec:background_modelprior} this amounts to adding additional record pairs directly to the ``u'' component, so we call this step a U-correction. We examine the effect of the U-correction further in Appendix~\ref{sec:ucorrection}.


\section{Linking the Great Registers: Alameda County}\label{sec:alameda}

In our study of the Great Registers we link 1932 and 1936 voter registration files for Alameda county. We chose this location and period because the data quality in Alameda is relatively good, prior work suggests that the party switching rate over this period is relatively high, and we suspect that registers from presidential election years have a higher degree of overlap than those from adjacent election years. 
Data preprocessing details appear in Appendix~\ref{sec:data-proc}. After cleaning and parsing the records from each year, we are left with 259,162 records from 1932 and 288,087 records from 1936. 

We present two estimates of the links between the 1932 voter file and the 1936 voter file.  The first are based on the Bayesian model described in  Section~\ref{sec:background_modelprior} with post-hoc blocking and restricted MCMC.  The second estimates are from a benchmark analysis using methods described in \cite{enamorado2018using}, as implemented in the fastLink R package \citep{fastLink_package}. FastLink utilizes a Fellegi-Sunter based model with parameter estimates computed via EM with specialized routines for blocking and post-estimation imposition of one-to-one matching constraints.  We present error rate comparisons between the estimated links in Section~\ref{sec:false_match_rates} and differences in estimated party switch rates in Section~\ref{sec:switch_rates}; further comparisons appear in supplemental material (Section~\ref{sec:fastlink-compariosns}).

These two estimates differ in both the initial indexing/blocking steps and the similarity cutoffs used to compute the comparison vectors, due to limitations in the fastLink package.  To isolate the effects of these choices, we also conducted a comparison between fastLink and a Bayesian model estimated using a set of record pairs and comparison vectors identical to those used by fastLink (Section~\ref{sec:comparable-model} in supplemental material). We found that the differences in results presented below are not primarily driven by the choice of indexing/blocking scheme or the selection of cutoffs in comparison vectors, but rather by specific modeling and estimation choices. We discuss these further in supplemental material (Section~\ref{sec:alameda_method_diff}).

\subsection{Bayesian Model: Implementation Details} \label{sec:alameda_bayesian}

Before estimating the Bayesian model we reduced the set of potential matches via indexing by disjunctions of blocking keys.  A record pair was included as a potential match if the first three characters of the given name or the first three characters of the surname matched exactly.\footnote{Women's first names beginning with ``mar'' are exceedingly common in this period, so for records pairs with both first names beginning with ``mar'' we used four characters of the first name.}  This left about 850 million record pairs as potential matches.

\subsubsection{Construction of comparison vectors} \label{sec:alameda_bayesian_comparison}

\begin{table}
    \centering
    \begin{tabular}{cc}
    \hline
    \hline
    Similarity Level & Similarity Range \\
    \hline
    6 & [1]\\
    5 & [0.85, 1)\\
    4 & [0.6, 0.85)\\
    3 & [0.45, 0.6)\\
    2 & [0.25, 0.45)\\
    1 & [0.0, 0.25)\\
    \hline
    \hline
    \end{tabular}\hspace{5.0em}%
    \begin{tabular}{cc}
    \hline
    \hline
    Similarity Level & Similarity Range \\
    \hline
    5 & [1]\\
    4 & [0.75, 1)\\
    3 & [0.5, 0.75)\\
    2 & [0.25, 0.5)\\
    1 & [0.0, 0.25)\\
    \hline
    \hline
    \end{tabular}
    \caption{String similarity to ordinal mapping.  Jaro-Winkler string similarity (left) and zero-padded Levenshtein string similarity (right).}
    \label{tab:ordinals}
\end{table}

We used the Jaro-Winkler similarity score  to compare first name, surname, occupation, and street name fields.  We compared the street number field with a Levenshtein distance, using zero-padding to ensure the distance is calculated between strings of equal length.  The resultant string similarities were converted to comparison vectors by binning (Table~\ref{tab:ordinals}). We compared our constructed female indicator and the street type using exact matching (coded as 2 for a match and 1 for a non-match).

Modeling the similarity in the middle name field required a more nuanced approach because in many cases only a middle initial was recorded. We considered three cases: two full middle names present, two middle initial initials present, and a full middle name present in one record and only a middle initial in the other record.  When two full names were present we used the same string comparison and cutoffs as for comparing first and surnames. For two initials, or one full name and one initial, we used exact matching between the first letter of the full name and the initial. This resulted in $10$ possible similarity levels for middle name, the six in Table~\ref{tab:ordinals} for two full middle names, two for comparisons between middle initials, and two for comparisons between a middle initial and a full middle name. 

\subsubsection{Prior distributions, post-hoc blocking and restricted MCMC} \label{sec:alameda_bayesian_estimate}

Our prior specification for the model in Section~\ref{sec:background_modelprior} began with a Beta-bipartite prior on $C$ with $\alpha = 1.0$ and $\beta = 1.0$. For first name, surname, occupation, and street name, $m_j\sim \mathrm{Dir}(10, 6, 2, 1, 1, 1)$.  For middle name $m_j\sim \mathrm{Dir}(10, 5, 3, 6, 2, 1, 1, 1, 1, 1)$, where the weights of 5 and 3 correspond to exact matching between two initials and exact matching between an initial and the first letter of a full middle name, respectively.  For street number $m_j\sim \mathrm{Dir}(10, 6, 2, 1, 1)$, and for female and street type we set $m_j\sim \mathrm{Dir}(5, 1)$.  The prior distribution for all $u_j$ vectors was uniform; these parameters are well-estimated from the data since most pairs are non-matches.

We fit the model via restricted MCMC using post-hoc blocks based on the maximal weights procedure detailed in Section~\ref{sec:assignment_maxweights} of the supplemental material. We set $N_c$, the maximum post-hoc block size, to 250,000 record pairs.  The resulting set of 94,997 distinct post-hoc blocks contains approximately 820,000 of the record pairs, a reduction of 99.9\%. 

On a 2014 Linux workstation, constructing the comparison vectors took about 2 hours, while estimating the weights and finding the post-hoc blocks took approximately 90 minutes. We ran the restricted MCMC algorithm for 25,000 steps, where each ``step'' comprises an update to the $m-$ and $u-$ paramaters in addition to a Metropolis-Hastings proposal within {\em every} post-hoc block (i.e., one step of the MCMC algorithm constitutes nearly 95,000 add/delete/swap updates to $C$, one within each post-hoc block). The restricted MCMC algorithm took 3.7 hours to run. In total it took under 7.5 hours to link these two files.

\subsection{fastLink implementation} \label{sec:alameda_fastlink}

We relied on internal fastLink functions to block on first name, since fastLink was unable to use indexing by disjunctions by design. We used fastLink's built-in clustering function to generate blocks of maximum size 10,000 by 10,000 -- much larger than our post-hoc blocks -- resulting in 123 distinct blocks containing 1.1 billion record pairs.  To generate comparisons with fastLink we used the Jaro-Winkler similarity score on all fields except for female and street type, for which we rely on exact matching.  We were limited to three categories for the string comparisons by the fastLink package, so we followed recommendations in \cite{winkler1990string,enamorado2018using} to set string similarity cutoffs; similarity above 0.92 corresponded to an ``exact'' match, between 0.88 and 0.92 a partial match, and below 0.88 a non-match on the field.\footnote{In very recent versions of the fastLink package the default 0.92 similarity threshold for an ``exact'' match was increased to 0.94.}.

\cite{enamorado2018using} suggest declaring matches by thresholding estimated posterior probabilities (derived from Eq.~\ref{eq:fsmix}) at a value between 0.75 and 0.95. Based on early inspection of the results we chose a cutoff of 0.9. Like other EM-based methods, fastLink does not incorporate a one-to-one matching constraint during estimation, so we relied on the fastLink's deduplication procedure to produce a set of record pairs consistent with the one-to-one matching assumption. Roughly, the fastLink deduplication procedure limits declared matches to those record pairs for which the estimated posterior probability is the observed maximum for both records across all possible record pairs involving one of two records.  In the case where multiple record pairs achieve the maximum, ties are broken by sampling uniformly among the candidates.

\subsection{Comparing the Bayesian model and fastLink}\label{sec:alameda_model_comparison}

Before estimating party switching rates we compared the links made by each method. In particular, we focus on estimating the overall match rate and the rate of false matches. Overall, we find that the Bayesian model makes many more true matches at a significantly lower false match rate. 

Since the fastLink software was not able to accommodate some of the choices we made in our Bayesian model -- specifically our approach to indexing and our use of more than three levels of string similarity -- we also fit a version of the Bayesian model that used the same string similarity cutoffs and blocking procedure as fastLink.  Results from the comparable model, presented in Section~\ref{sec:comparable-model} of the supplemental material, are very similar to our preferred Bayesian model. We conclude that the differences noted below are not due to indexing/blocking choices or differences in string similarity cutoffs. In Section~\ref{sec:alameda_method_diff} of the supplemental material we discuss other possible explanations for our findings below.

\subsubsection{Estimating False Match Rates}\label{sec:false_match_rates}

We compared the false match rate of fastLink and Bayesian estimators by manually confirming matches declared by one or both methods, blind to which method actually made the match. We pre-registered the design of this comparison \footnote{Available at http://egap.org/registration/5452.}. For this exercise we needed to reduce the full Bayesian posterior over linkage structures to one set of declared links. We used a simple point estimate, classifying any record pair with a posterior match probability of greater than 0.5 as a match. This is the Bayes estimate under squared error or balanced misclassification loss functions \citep{tancredi2011hierarchical}. For the fastLink estimate we use a conservative threshold of 0.9, which is within the recommended range of 0.75 to 0.95 \citep{enamorado2018using}.  These two thresholds are not directly comparable -- fastLink's estimated matching probabilities are conditional on the block and are not proper posterior probabilities, since they fail to enforce the one-to-one constraint and often sum to values larger than one. By contrast, the posterior matching probabilities from our Bayesian model are proper probabilities and are not conditional on the indexing scheme due to our U-correction. We begin by comparing the point estimates, which guided our hand labeling, before turning to a more apples-to-apples comparison in the next subsection.

The two estimated sets of matches define three disjoint sets: record pairs classified as a match by both models; Bayesian only matches (record pairs classified as a match by our Bayesian model but not by fastLink); and fastLink only matches (record pairs classified as a match by fastLink but not by our Bayesian model).  We further subdivided these sets of matches into two strata: ``mover'' and ``non-mover'' matches. We suspected that links made between individuals at different addresses would have higher error rates regardless of method. We defined a matched record pair as a ``mover'' if the string similarity between the street names was less than 0.85 or the similarity between the street numbers was less than 0.5\footnote{These similarity thresholds correspond to a similarity level of less than 5 for street name and less than 3 for street number as listed in Table~\ref{tab:ordinals}.}. Otherwise the match was classified as a non-mover, including cases where the address information is missing for one or both records.

For both mover and non-mover matches we drew a stratified sample, sampling 100 matches from the intersection stratum and 150 matches from each of the Bayeisan only and fastLink only strata. This yielded 800 total pairs for labeling (400 mover matches and 400 non-mover matches). Each record pair was then labeled as either a false match (FM), a true match (TM), or no determination (ND) (for record pairs where there was not enough information to classify the record pair as either a match or a non-match with a reasonable level of confidence).  The labeller was presented with the record pair under question, as well as similar records from each file, but was blind as to which method(s) had made the match.  Results of the labeling are shown in Table~\ref{tab:labeling}.

\begin{table}
\centering
\resizebox{0.99\textwidth}{!}{%
\begin{tabular}{l|rrrrr|rrrrr|rrrrr}
\hline
\hline
\multicolumn{1}{c}{} & \multicolumn{5}{c}{Mover} & \multicolumn{5}{c}{Non-Mover} & \multicolumn{5}{c}{Overall} \\
Stratum & FM & TM  & ND & Labeled & Total Matches & FM & TM  & ND & Labeled & Total Matches & FM  & TM  & ND & Labeled & Total Matches \\
\hline
Intersection  &   2 &  88 &  10 & 100 & 14,276 &   4 &  96 &   0 & 100 & 38,968 &   6 & 184 &  10 & 200 & 53,244\\
Bayesian Only &  12 & 118 &  20 & 150 & 18,525 &  26 & 121 &   3 & 150 & 60,562 &  38 & 239 &  23 & 300 & 79,087\\
fastLink Only & 102 &  33 &  15 & 150 & 21,636 & 105 &  44 &   1 & 150 &  4,348 & 207 &  77 &  16 & 300 & 25,984 \\
\hline
\hline
\end{tabular}
}
\caption{Hand-coding results from mover (left) and non-mover (center) matches and overall (right).  Each matched record pair is labeled as either a false match (FM), a true matches (TM) or no determination (ND), when insufficient information is available.}
\label{tab:labeling}
\end{table}

Our pre-registered comparison excluded matches labeled ND. Removing these pairs we estimated the mover and non-mover false match rates for both our Bayesian model and for fastLink, computing estimates and standard errors using standard methods for stratified samples \citep{rice2006mathematical}. These appear in Table~\ref{tab:model_fmr_estimates}. The overall estimated false match rate is 0.11 for the Bayesian model and 0.27  for fastLink, a difference of 0.16 ($p<0.0001$). The difference is driven primarily by fastLink's high error rate in the mover stratum (a difference of 0.40 $\pm0.05$, $p<0.0001$). For non-movers, the difference between the two methods was not statistically significant, and the Bayes estimate captured many more matches (Table~\ref{tab:labeling}).

A more conservative approach would count all the ``ND'' labeled record pairs as false matches.  While this is likely to overestimate the true false match rate, since some ND matches correspond to true matches, it furnishes something of an upper bound for the true false match rate.  We repeated the analysis counting ND record pairs as false matches.  This gives higher estimated false match rates across the board, but results in the same conclusions as when ND record pairs are excluded from the analysis (Table~\ref{tab:model_fmr_estimates}).

\begin{table}
\centering
\resizebox{0.95\textwidth}{!}{%
\begin{tabular}{l|ccc|ccc}
\hline
\hline
\multicolumn{1}{c}{} & \multicolumn{3}{c}{ND Excluded} & \multicolumn{3}{c}{ND as Non-match} \\
 & Mover & Non-Mover & Overall & Mover & Non-Mover & Overall \\
\hline
Bayesian Model & 0.062 (0.031, 0.093) & 0.123 (0.083, 0.164) & 0.108 (0.077, 0.139) & 0.173 (0.126, 0.219) & 0.133 (0.092, 0.175) & 0.143 (0.110, 0.176) \\
fastLink & 0.464 (0.419, 0.509) & 0.107 (0.071, 0.142) & 0.269 (0.240, 0.297) & 0.518 (0.470, 0.565) & 0.107 (0.072, 0.142) & 0.293 (0.264, 0.322) \\
\hline
Absolute Difference & 0.402 (p$<$0.0001) & 0.017 (p = 0.45) & 0.161 (p$<$0.0001) & 0.345 (p$<$0.0001) & 0.026 (p = 0.24) & 0.150 (p$<$0.0001) \\ 
\hline
\hline
\end{tabular}
}
\caption{Estimated false match rates with 95\% confidence intervals excluding ND record pairs (left) and counting ND record pairs as false matches (right) by model.}

\label{tab:model_fmr_estimates}
\end{table}

\begin{table}
\centering
\resizebox{0.99\textwidth}{!}{%
\begin{tabular}{l|ccc|ccc}
\hline
\hline
\multicolumn{1}{c}{} & \multicolumn{3}{c}{ND Excluded} & \multicolumn{3}{c}{ND as Non-match} \\
      & Mover & Non-Mover & Overall & Mover & Non-Mover & Overall \\
\hline
Intersection & 0.022 (0.000, 0.053) & 0.040 (0.002, 0.078) & 0.035 (0.006, 0.065) & 0.120 (0.056, 0.184) & 0.040 (0.002, 0.078) & 0.061 (0.029, 0.094) \\ 
Bayesian Only & 0.092 (0.043, 0.142) & 0.177 (0.115, 0.239) & 0.157 (0.108, 0.206) & 0.213 (0.148, 0.279) & 0.193 (0.130, 0.257) & 0.198 (0.147, 0.249) \\ 
fastLink Only & 0.756 (0.683, 0.828) & 0.705 (0.631, 0.778) & 0.747 (0.685, 0.809) & 0.780 (0.714, 0.846) & 0.707 (0.634, 0.780) & 0.768 (0.711, 0.824) \\  
\hline
\hline
\end{tabular}
}
\caption{Estimated false match rates with 95\% confidence intervals excluding ND record pairs (left) and counting ND record pairs as non-matches (right) by stratum.}
\label{tab:strata_fmr_estimates}
\end{table}

Examining the estimated false match rates for the individual strata, as shown in Table~\ref{tab:strata_fmr_estimates}, helps to explain the differences in error rates.  As expected, the estimated false match rate in the intersection stratum is  lower than the Bayesian only or fastLink only strata for both movers and non-movers.  However, the false match rate in the fastLink only stratum is much larger than in the Bayesian only stratum. In fact, for both mover and non-mover matches, we estimate that the \textit{majority} of the matches in the fastLink only strata are false matches. In contrast, the estimated overall error rate in the Bayesian only stratum is nearly five times lower than the fastLink only stratum when excluding ND pairs ($p<0.0001$), and over four times lower when counting ND pairs as true non-matches ($p<0.0001$).

\subsubsection{False Match Rates as a Function of Linked Sample Size}
\begin{figure}
  \centering
  \includegraphics[scale=1.0]{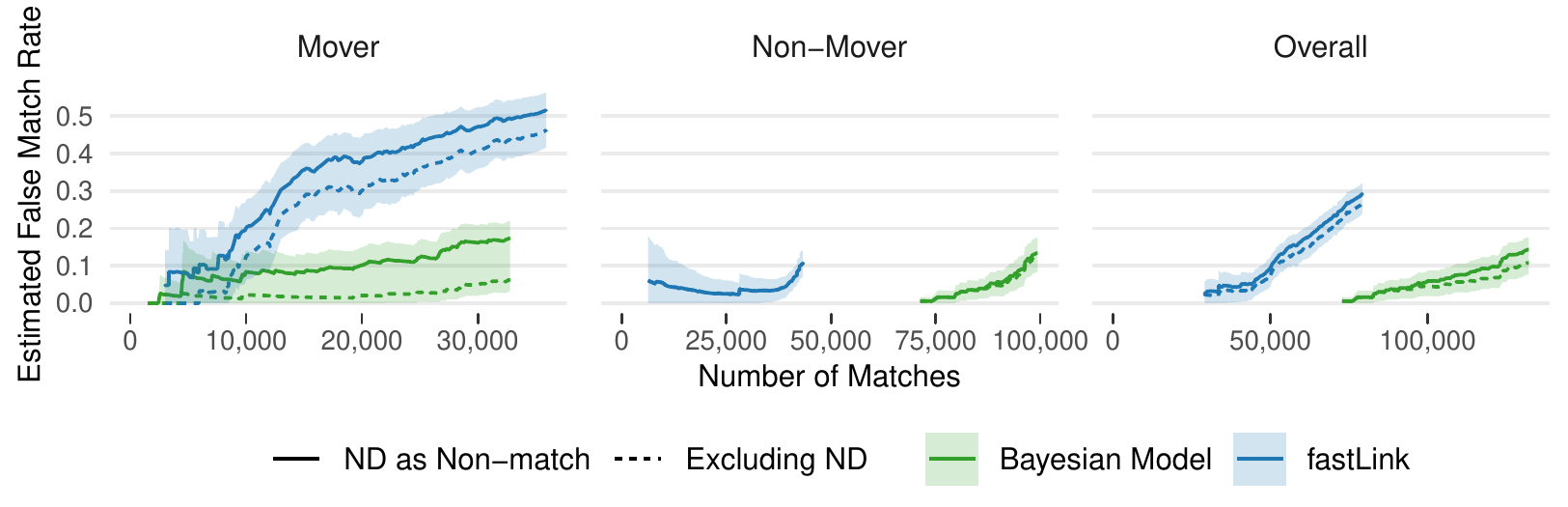}
  \caption{Estimated false match rates for mover matches (left), non-mover matches (center), and all matches (right) for deduplicated fastLink (blue) and Bayesian model (green). Solid lines count ND (``no determination'') record pairs as true non-matches, dashed lines exclude such pairs. Bands are the union of 95\% confidence intervals counting ND pairs as non-match and excluding ND pairs.}
  \label{fig:fmr}
\end{figure}

One limitation of the estimates reported in Table~\ref{tab:strata_fmr_estimates} is that they correspond to only a single threshold for each model, and for reasons described above these two thresholds are not directly comparable. This is a common problem when evaluating record linkage methods; to mitigate this issue \cite{hand2018note} suggest comparing methods by plotting error rates or other performance metrics as a function of the number of matches made as thresholds vary. Figure~\ref{fig:fmr} shows these curves for the overall false match rate and the false math rate in mover/non-mover strata. Overall the Bayesian method makes significantly more matches than fastLink at any given false match rate. This is due mostly to the higher match rate in the non-mover stratum, although a similar pattern can be observed in the mover stratum.

\section{Party Switching Rates in Alameda County}\label{sec:switch_rates}

Political scientists have postulated that the conversion of Republicans into Democrats was led by working class voters and women, arguing that more members of these groups switched parties than other segments of the electorate \citep{corder2016counting, sundquist1983dynamics}. That such a change occurred is readily apparent in the Great Registers \citep{spahn2018before}.   
Based on cross-sectional data, before the realignment (in 1928) all demographic groups had a Democratic registration rate of about 20\%   \citep{spahn2018before}. This rate rose significantly during the realignment period, indicating that most party switches were from the Republicans to the Democrats. Among men registered with one of the two major parties, blue collar men were 21 percentage points more likely to be registered as a Democrat in 1936 than in 1932. Among white collar men, the change was just 14 points. In Alameda county, women and men moved about the same amount, each increasing their support for the Democrats by a bit less than 20 percentage points. 

However, cross-sectional data can only tell a limited story about party switching --  for example, it is unable to disentangle the effects of party switching from differential turnout. A more nuanced picture can be drawn if individuals can be linked over time. Since name prefix and occupation are identified for individuals in the Great Registers, and gender can be inferred with a fairly high degree of confidence, given a set of links it is easy to estimate both the cross-sectional partisan composition and the party-switching rates for individuals registered in both years and aggregate up to groups in order to shed light on these theories.  

In this section we consider links generated by our Bayesian model and fastLink. To simplify the presentation of results, voters that were registered as neither Democrat nor Republican (10.4\% in 1932 and 7.5\% in 1936) in either of the two election years are excluded from the analysis in this section. The largest possible number of links between such individuals is 232,106 (the number of voters with a major party affiliation in the smaller 1932 register). In reality this is a very loose upper bound, as it would only be attainable if everyone in registered in 1932 with a major party affiliation also registered in 1936 in the same county (i.e. no drop-out, death, or out-migration).

Fitting the Bayesian model produced a posterior mean of about 117,000 individuals linked across the two files with major party affiliation in both years. FastLink returned fewer links, identifying around 69,000 record pairs as matches.   The composition of the linked sets were also markedly different; for example, both methods linked approximately the same {\em number} of movers, so the proportion of fastLink's matches that are movers is about twice as high. 

\subsection{Party Switching Rates}\label{sec:switch_rates_party}

We estimated the posterior distribution of party-switching rates from the Bayesian model by computing them for every MCMC sample of $C$, after discarding the first 2,500 iterations as burn-in. For fastLink we obtain a set of links by thresholding its estimated link probability at $0.9$. \cite{enamorado2018using} recommends weighting various estimates computed over linked data by its estimated match probabilities to account for linkage uncertainty.  However, when estimating the party switching rate it is necessary to account for uncertainty both in the individual links and in the total number of links.  This setting does not seem to fit in the cases considered in \cite{enamorado2018using}, and we are unaware of an adjustment that appropriately characterizes linkage uncertainty here using only the output provided by fastLink. Moreover, the fastLink estimated probabilities don't seem to be well-calibrated posterior probabilities (see Table~\ref{tab:example_posterior_comparison} for an example), so any adjustment based on them is perhaps questionable. For the purposes of comparing the methods here we present only point estimates using the fastLink matches.

We also considered adjusting point estimates based on the estimated false match rate. We expect erroneous matches to inflate the estimated match rate: Even over this period party switching is not the norm, so false matches are more likely to show a switch in parties than true matches. Considering the distribution of party affiliations in the two files linking two records at random will show a party-switch about half the time. Assuming false matches occur completely at random, the set of estimated matches is composed of a mixture of false matches with proportion $\pi_F$, the false match rate, and true matches with proportion $(1-\pi_F)$, which have a switching rate of $\rho_T$ (our target of inference). The observed switch rate is related to $\rho_T$ by
\[\rho_{observed} = 0.5\pi_F + \rho_T(1-\pi_F).
\]
Using this formula and an estimate of the false match rate we can convert the observed switch rate to an estimate of the true switch rate using the relationship 
\[
\rho_T = \frac{\rho_{observed} - 0.5\pi_F}{(1-\pi_F)}.
\]

Using the labeled data from Section~\ref{sec:false_match_rates} we estimate false match rates both overall and within subgroups (adjusting the stratum proportions to match those of the subgroups in the latter case). For the Bayesian model we apply this adjustment to each MCMC sample to obtain the posterior of the adjusted switching rates. We consider two estimates of the false match rate: ``Adjusted'' estimates use a false match rate in which ND labeled pairs are excluded while the ``Adjust+'' estimates count ND labels as true non-matches (errors)\footnote{In estimating the false match rate for the Bayesian model record pairs falling into none of the strata sampled for labeling, because they were assigned a low posterior link probability by every algorithm, are assumed to have either the maximum estimated false match rate across strata (Adjusted) or a false match rate of 1 (Adjust+).}. We suspect that these error rates under- and over-estimate the true error rates respectively (for reasons discussed in Section~\ref{sec:false_match_rates}), so it seems reasonable to regard them as providing sensitivity bounds for the true party switching rate.

\begin{figure}
  \centering
    \includegraphics[width=0.9\linewidth]{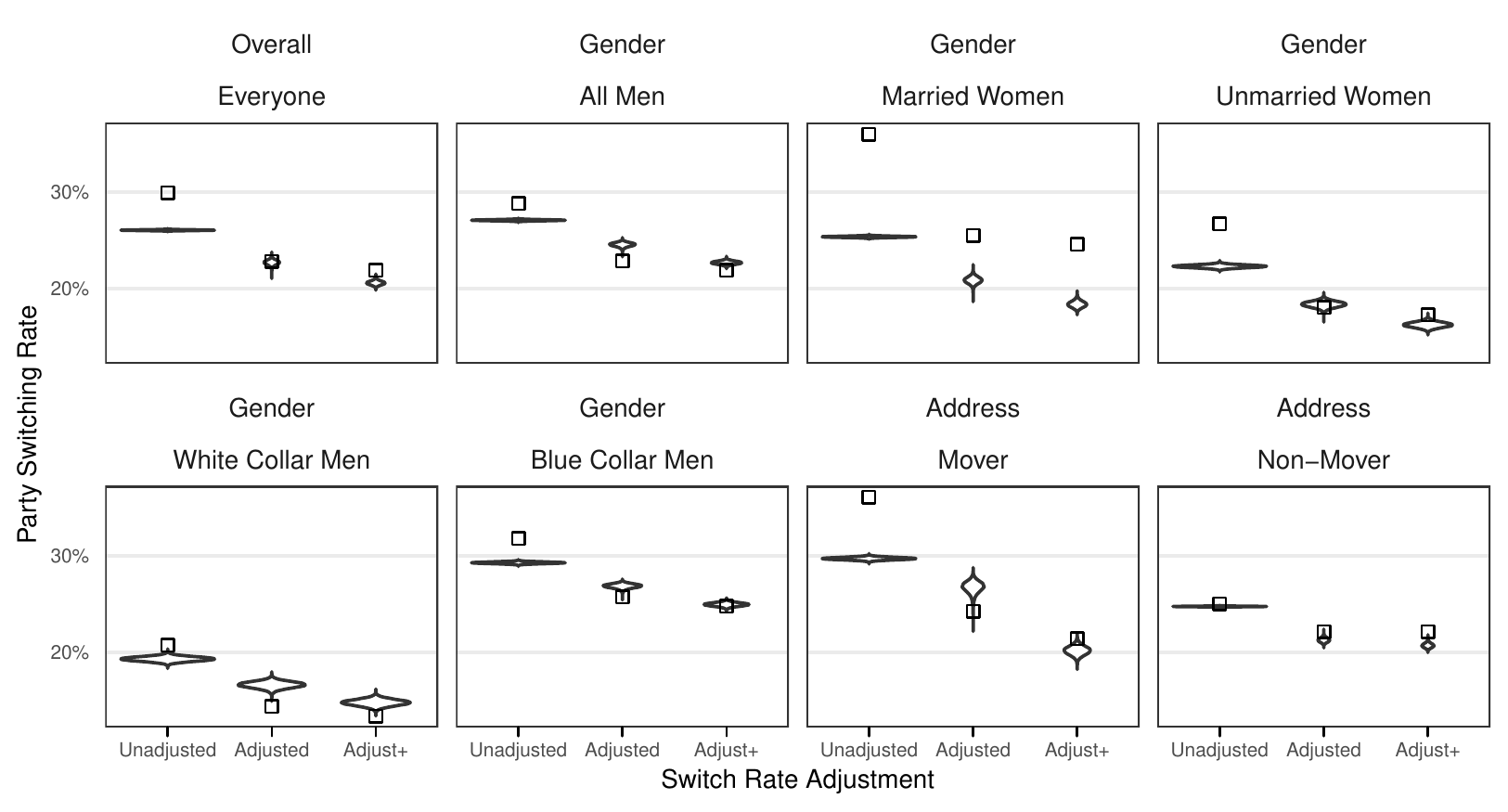}
  \caption{Posterior distributions of party switching rate for interesting subgroups across samples of record-pairs for the Bayesian model. The square points show the point estimates from fastLink. The bias-adjusted switch rate (``Adjusted'') and the bias-adjusted switch rate treating indeterminate matches as false matches (``Adjust+'') are also plotted. }
  \label{fig:switching_by_subgroup}
\end{figure}

The distribution of the mean party-switching rate overall and for interesting subgroups is displayed in Figure~\ref{fig:switching_by_subgroup}.  With the exception of non-movers, fastLink consistently shows higher unadjusted rates of party-switching than the Bayesian model, save for the overall non-mover stratum where the two are nearly identical. This is consistent with our findings in Section~\ref{sec:false_match_rates}, where fastLink's higher overall false match rates were driven largely by its behavior in the mover stratum.  

After adjustment most of the estimates exhibit better agreement between fastLink and the Bayesian model. So why do we prefer the Bayesian model over fastLink? First, the Bayesian method returns a much larger matched set -- about 70\% larger, or 48,000 links -- with lower overall false match rate, and provides uncertainty intervals via the posterior distribution. Second, the estimates are not always brought in line by adjustments, for example in the case of married women in Fig~\ref{fig:switching_by_subgroup}. It seems reasonable to put more trust in the Bayesian model with its lover overall false match rate here, especially because it estimates a lower switching rate (consistent with making fewer false matches).  Third, the adjustment is imperfect. It assumes that false matches occur at random and that the only variation in false match rates by demographic subgroup is due to the varying proportions of movers/non-movers within that group, both of which are likely oversimplifications. Relaxing these assumptions would require at minimum a much more extensive labeling exercise. Therefore we recommend choosing a method with low overall estimated false match rates, and then applying false match rate adjustments as feasible, treating this as a form of sensitivity analysis. We do not consider estimates from fastLink any further.

\begin{figure}
  \centering
    \includegraphics[width=0.9\linewidth]{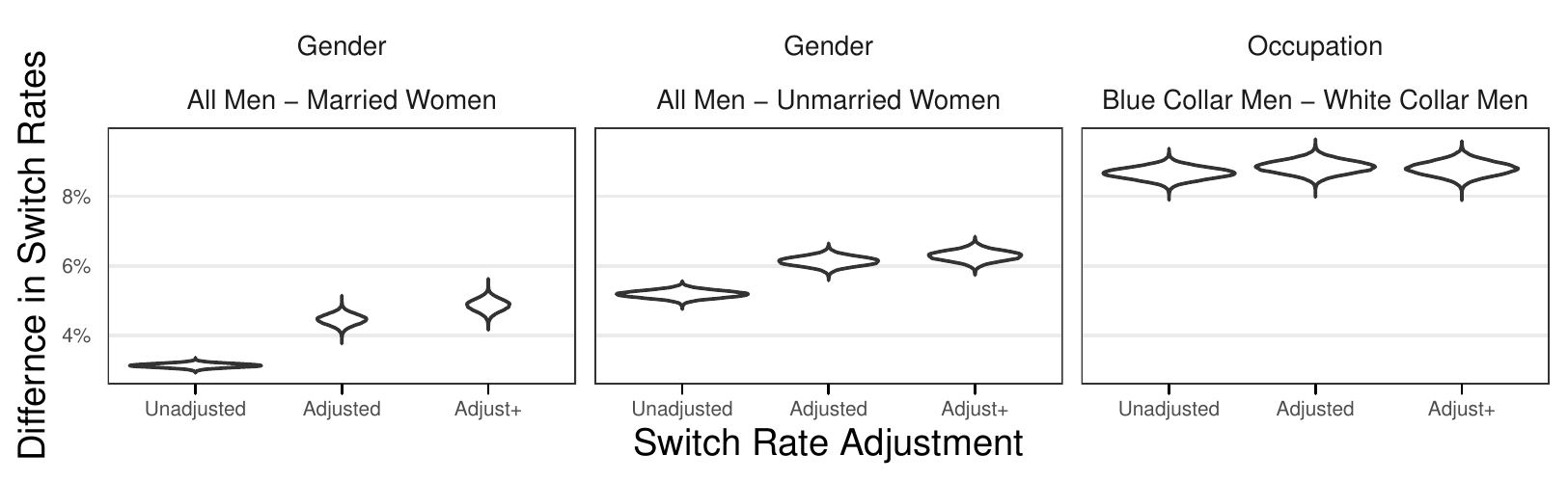}
  \caption{Posterior distribution of the difference in mean switch rates between subgroups. Posteriors of the bias-adjusted switch rate (``Adjusted'') and the bias-adjusted switch rate treating indeterminate matches as false matches (``Adjust+'') are also plotted.}
  \label{fig:switch_rate_diff}
\end{figure}

\subsubsection{Demographic differences in party switching rates}

To draw out the differences in switch rates based on gender and class we plot the difference in various switch rates in Figure~\ref{fig:switch_rate_diff}. The rightmost panel shows a stark difference in party switching rate between blue and white collar men (the rate is estimated at about 8.5\%, regardless of bias adjustment). The high switching rates among blue collar voters confirm what's long been known: that blue collar workers led the realignment towards Roosevelt's Democratic party. This particular fact has been known at the group level \citep{ladd1975transformations}, but has never been demonstrated with individual-level data.

Separating the political attitudes of men and women is considerably harder because they tend to be clustered together in space, voting in the same places. Though Corder \& Wolbrecht \citeyearpar{corder2016counting} use ecological inference methods to try to separate the political behavior of men and women, such approaches will always prove difficult because of low variation in the gender ratio. Individual-level data is much better suited to the task. The leftmost panels of Figure~\ref{fig:switch_rate_diff} show that men switched parties at a considerably higher rate than women, which at first glance seems contrary to what one would expect based on Corder \& Wolbrecht's analysis. We return to this in the next subsection, where we disaggregate total party switching into flows to and from Democratic party, providing a slightly more nuanced picture.

\subsubsection{Relative Party Switching Rates}\label{sec:switch_rates_party_fraction}

\begin{figure}
  \centering
    \includegraphics[width=0.9\linewidth]{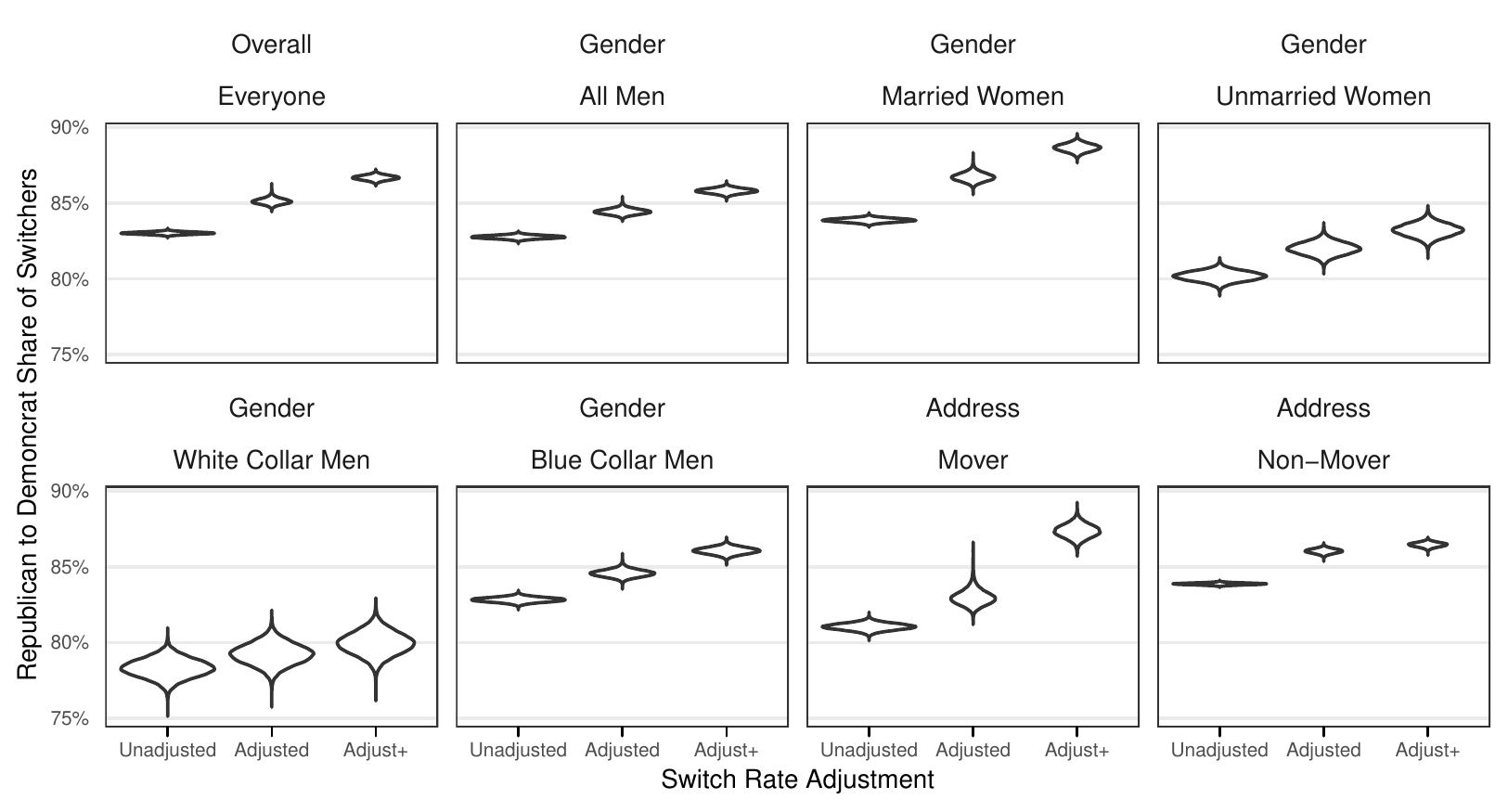}
  \caption{Posterior distributions of the fraction of party switchers who switch from the Republican party to the Democratic party. The overwhelming move towards to Democratic party is readily apparent among all subgroups but especially among married women and blue collar men.}
  \label{fig:switch_rate_direction}
\end{figure}

A high party switching rate does not in and of itself guarantee a large flow from one party to the other, as it could be the case that large numbers of voters are switching parties but the flows roughly cancel in aggregate.  Figure~\ref{fig:switch_rate_direction} shows the fraction estimated fraction of party switchers that switch from the Republican party to the Democratic party, indicating that the vast majority of switching was from Republicans to Democrats, as expected.

As in the previous section we adjusted these estimates for potential false matches. Given the share of voters registered to the Democratic and Republican parties in 1932 and 1936, randomly linked record pairs which differ on party would appear to switch from the Republican party to the Democratic party approximately 70\% of the time. This proportion is nearly identical when computed separately for men and women.  We modify the adjustment introduced in Section~\ref{sec:switch_rates_party} and compute the adjusted fraction as $\tau_{adj} = (n_{R2D} - 0.7n_{switch}\pi_{F}) / n_{switch}(1 - \pi_{F})$ where $n_{R2D}$ is the observed number of voters switching from the Republican to the Democratic party, $n_{switch}$ is the observed number of voters switching party, and $\pi_{F}$ is the false match rate.  The false match rates used for the``Adjusted'' and ``Adjusted+'' estimates remain the same.

While Figure~\ref{fig:switch_rate_diff} shows that men switched parties at a considerably higher rate than women we see a more mixed picture in Figure~\ref{fig:switch_rate_direction}.  While all groups switch to the Democratic party at a much higher rate than the Republican part, the group with the most lopsided switching is married women (followed by blue collar men). Thus, Figure~\ref{fig:switch_rate_direction} offers a picture more consistent with the findings of  Corder \& Wolbrecht's: Unmarried women appear to favor the Democratic party less than any other group except for white collar men. 

Though it's hard to say for sure why unmarried women (who are presumably younger) would have realigned less than their married women counterparts, one possibility is that they were more influenced by the parents who are older (and, on average, more Republican) than married women's spouses, who are closer to the same age. The gap is not explained by different base rates of Democratic affiliation -- in 1932, the unmarried and married women posterior mean rate of Democratic registration of 26.5\% and 26.2\%, respectively. 

Of course, it may also be the case that our linked sample is somehow less representative for unmarried women than for other groups -- in general, over this time period women are more difficult to link than men. However, for the estimands considered here, this differential non-linkage is only consequential if non-linkage {\em within} the stratum of unmarried women is correlated with party-switching. (This is why we focus on estimating rates rather than totals.) This seems relatively unlikely, although not impossible. 
Finally, we note that our analysis cannot provide a complete picture of party affiliation for women over this period, as our analysis of women excludes those who marry between 1932 and 1936. The vast majority of these women would have changed their name and moved to a new address, and they are simply unlinkable using only the Great Registers.   

To further understand the implications of potential differential false non-match rates, we  compared the distribution of key demographics in 1932 and 1936 for the linked subsample to the distribution across the entire register in each year (Appendix~\ref{sec:linked-comp}).  The distributions are rather close, particularly for marital status of women and occupation of men.  Some observed differences are expected; our linked sample skews slightly male, likely due in part to reasons discussed above. Indeed, while we expect these distributions to be grossly similar, we would not expect all of them to match exactly even if linkage was done perfectly because of differential voter turnout (including drop-out and new registrations). For example, based on the increase in the total number of voters in 1936 and the increase in Democratic registration over 1932 we might expect the distribution of new registrants in 1936 to skew Democratic, which would tend to make a perfectly linked subsample (all of whom were registered in 1932) proportionally more Republican than the registered population in 1936. In fact, our linked sample skews slightly Republican by three percentage points in 1932 and five in 1936 (possibly due to compound effects of differential turnout and Republicans being on average older, wealthier, and more stable, and therefore easier to link).

\section{Discussion and Future Work}\label{sec:discussion}

Bayesian probabilistic record linkage models provide an appealing framework for performing record linkage: They can provide accurate point estimates of links between records, and they allow for uncertainty in the links between to be quantified and propagated through to subsequent inference.  The main barrier to their adoption in practice has been computational. Post-hoc blocking and restricted MCMC make Bayesian modeling for PRL feasible for much larger problems, as demonstrated by our analysis of the Great Registers. 

Linking our extract from the Great Registers provides new insight into party switching over one of the largest political realignments in American history. However, these insights are necessarily limited by the information available on the registration rolls. In particular we are missing important demographics like age and education, and variables like gender, marital status, and occupation type have to be reconstructed from attributes available on the file. It may be possible to obtain a more fine-grained picture of party switching by linking the registers to adjacent Census years to pick up more detailed demographic information about voters, provided these additional links can be made accurately. Similarly, using additional data sources like marriage records to help bridge the gap between election years could reduce error rates for difficult subpopulations, such as women who marry during the ``off'' years. 

Clearly it would be useful to adapt our methods to link multiple files simultaneously, both for linking in the Great Registers and for many other applied problems.  The computational challenges inherent in record linkage grow rapidly as we consider linking multiple files; see e.g. \citet{sadinle2013bayesian,steorts2015entity,steorts2016bayesian} for some examples and discussion of model-based deduplication and multiple file linking. Adapting post-hoc blocking to the multiple file setting (and to files with duplicates) is a promising area for future work. We expect that the simplest multiple-file/de-duplication version of post-hoc blocking -- stacking the multiple files and proceeding as in Section~\ref{sec:posthoc} as though we were matching this ``meta'' file to itself -- should work well, provided adequate post-hoc blocking weights can be constructed.

\bibliographystyle{apa}
\bibliography{recordlinkage,assignment,misc,realignment}
\newpage
\appendix

\section{Data processing details}\label{sec:data-proc}
Before constructing comparison vectors we undertake a number of pre-processing steps. The suffix field is coded as missing so frequently that we are forced to discard it entirely.  Name prefix is also largely missing but is useful in that the vast majority of non-missing entries are either ``mrs", ``ms", or ``miss" indicating that the individual is a woman, a feature which is not coded explicitly in the original data.  We construct an indicator variable for probable females if one of these prefixes\textbf{} appears, or if the occupation is recorded as ``housewife'' or a variant thereof.  We then code the occupation variable as missing for housewives, as chance agreement on occupation for housewives is very common.  

We split the given name field into separate first name and middle name fields.  We also split the address field into three parts: street number, street name, and street type.  Street number is coded as missing in cases where the street number is not included in the address.  The street type was re-coded (e.g. mapping both ``rd'' and ``road'' to ``road'') to standardize common abbreviations. The street name contains the remains of the original address field after removing the street number and street type from the original address string. We discarded records missing two or more of the first name, surname, occupation, and street name fields.

\section{The impact of the ``U-correction'' on Bayesian models}\label{sec:ucorrection}

It is well known that failing to account for the blocking or indexing scheme can result in bias in the estimated $u$-parameters \citep{murray2016probabilistic}. This bias is introduced because, under any reasonable indexing or blocking scheme, the distribution of even non-matching comparison vectors will differ substantially between record pars within the scheme and those excluded from it. For example, in the indexing scheme employed for our Bayesian model (Section~\ref{sec:alameda_bayesian}) we consider only record pairs which match on the first three digits of first or last name. By design the scheme excludes record pairs which are dissimilar on both first name and last name, the overwhelming majority of record pairs.  This means that record pairs displaying low levels of similarity on first name or last name will be underrepresented, potentially massively so, relative to what would be observed if all comparison vectors were computed.  

Setting $C_{ab}=0$ for all record pairs outside of the blocking scheme then, under a conditional independence assumption, the missing comparisons need only be generated marginally.  That is, it is not necessary to calculate the full comparison vector for each record pair outside of the indexing or blocking scheme, only to determine frequency with which each similarity level would occur for each feature.  This is a result of the fact that, under a conditional independence assumption, the likelihood factors in such a way that only the marginal frequencies with each group (match and non-match) are necessary as can be seen in \eqref{eq:condindep}.

Computing these marginal frequencies is much more tractable as it can be done by computing similarities only for observed unique values and weighting appropriately. For example, the first name john occurs 8,173 times in the 1932 data and the first name william occurs 8,349 times in the 1936.  Hence, there will be a total of 68,236,377 record pairs (8,173$\times$8,349) in which the first name in 1932 is john and the first name in 1936 is william but the string similarity need only be computed once.  

\begin{table}
    \centering
    \begin{tabular}{lrrr}
    \hline
    \hline
    Field & Unique Values 1932 & Unique Values 1936 & Required Comparisons\\
    \hline
    First Name & 16,496 & 15,045 & 248,182,320 \\
    Middle Name & 6,489 & 6,039 & 39,187,071 \\
    Middle Initial & 37 & 32 & 1,184 \\
    Surname & 55,648 & 54,715 & 3,044,780,320 \\
    Female & 2 & 2 & 4 \\
    Occupation & 17,164 & 18,115 & 310,925,860 \\
    Street Number & 9,071 & 10,098 &  91,598,958 \\
    Street Name & 15,256 & 8,316 & 126,868,896 \\
    Street Type & 12 & 12 & 144 \\
    \hline
    \hline
    \end{tabular}
    \caption{Unique values observed within each year for each record field. Due to ORC errors numbers are sometimes observed in the middle initial field causing the number of unique observed values to be greater than 26.}
    \label{tab:unique_values}
\end{table}

In Table~\ref{tab:unique_values} we show the number of unique values observed for each field for each year. In total computing the full set of comparisons requires approximately 5 billion different comparisons. We note that while this is larger than the number of comparison vectors we generate for the analysis because multiple comparisons are required for each comparison vector even computing the full set of comparisons is less computationally costly than generating the comparison vectors. Furthermore,  this is less than 1\% of the nearly 675 billion comparisons that would be required if all possible comparison vectors were generated individually.  While significantly more efficient approaches exist for generating the full set of comparison vectors (e.g. the implementation in \cite{enamorado2018using} using a hash table), at large scales other constraints, such as memory constraints, often become binding.  Finally, while in our case we compute the marginal frequencies exactly it is possible to approximate them closely by computing similarities for record pairs generated via weighted random sampling of the unique values (with weights corresponding to observed frequencies). We refer to the inclusion of these frequencies as making a ``U-correction'' since the main effect is on the estimates of the $u$-parameters.  We find that this correction is extremely important in practice. 

\begin{figure}
  \centering
  \includegraphics[scale=1.0]{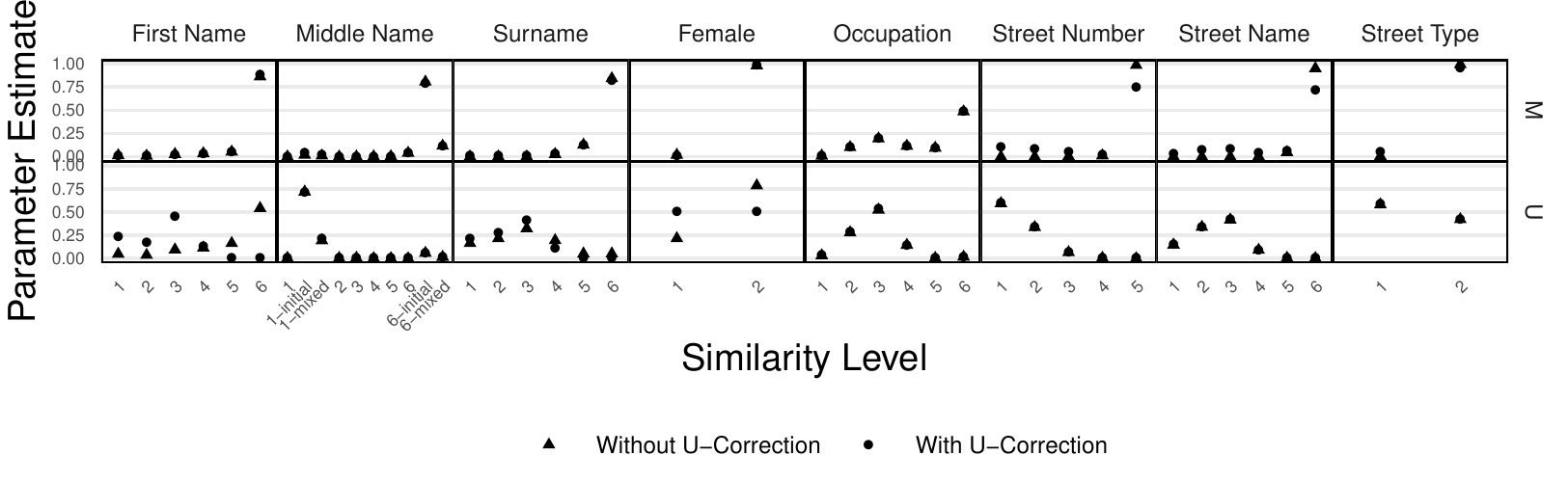}
  \caption{Posterior means of $m$-parameters (top) and $u$-parameters (bottom) with and without the U-correction.}
  \label{fig:parameters_preferred_model}
\end{figure}

We show the effect making the U-correction has on the parameter estimates in Figure~\ref{fig:parameters_preferred_model}.  Examining the $u$-parameters in the bottom row it is clear the largest effect is on first name, one of the indexing fields, and the female indicator, which is likely correlated with first name.  We see a smaller effect on surname, the other indexing field.  Interestingly we also see changes in the estimated $m$-parameters for the address fields street number, street name, and street type.  Here the observed change is a \textit{decrease} in the estimated probability of observing an exact agreement in the address fields, conditional on the record pair corresponding to a match.  This suggests that the model with the U-correction is matching a larger number of record pairs which differ on address, corresponding to the mover category discussed in Sections~\ref{sec:alameda_model_comparison} and \ref{sec:switch_rates_party}.  The differences in parameter estimates also results in a larger number of estimated links.

\begin{figure}
  \centering
  \includegraphics[scale=1.0]{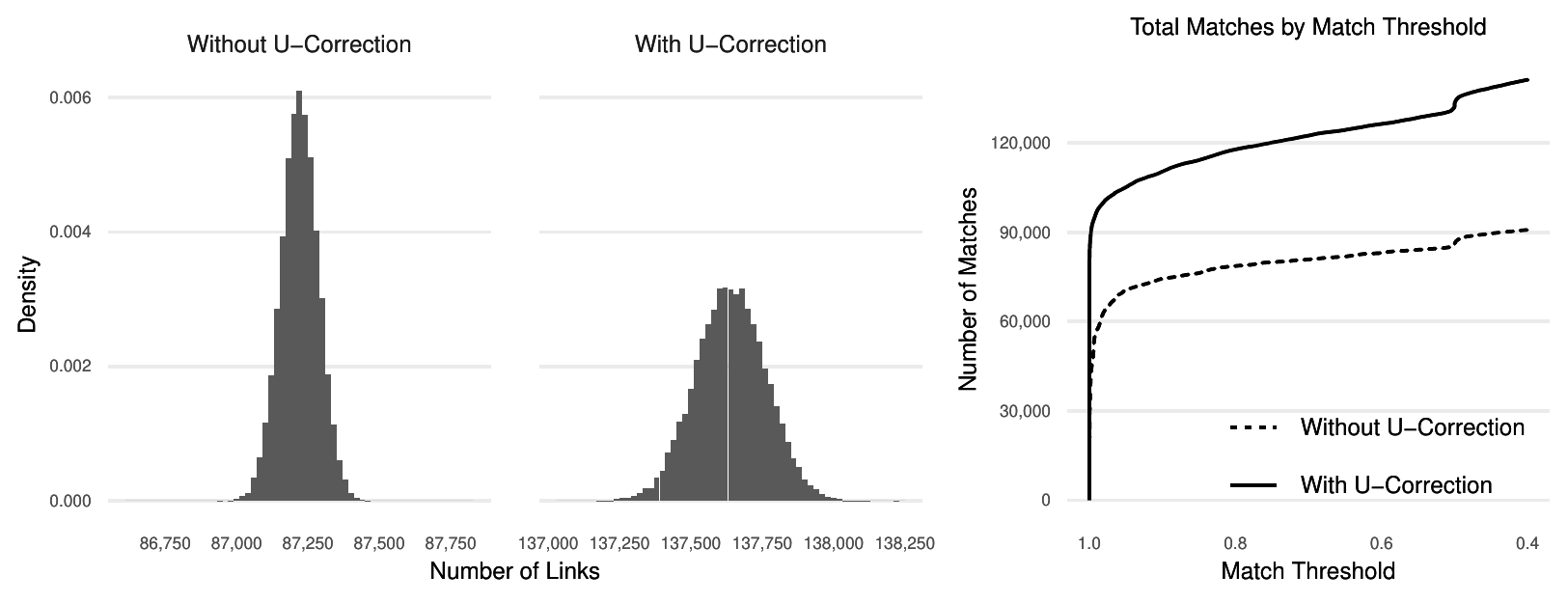}
  \caption{Posterior distribution of number of links for our Bayesian model with and without U-correction (left).}
  \label{fig:ucorr}
\end{figure}

Posterior distributions for the number of links with and without the U-correction are shown in Figure~\ref{fig:ucorr}.  The posterior distribution for the model with the U-correction (center) indicates both a larger mean number of estimated links, approximately 137,600 with the correction and only 87,200 without, as well as more dispersion in the number of links.  The right panel shows the number of matches that would be returned by each model as a function of the posterior probability threshold for declaring a record pair a match.  In addition to identifying more matches, for all match thresholds, the larger slope of the U-correction line indicates that the model identifies many more record pairs with a non-trivial level of uncertainty about the match status, perhaps better characterizing the matching uncertainty. An inspection of a set of links assigned a substantially high link probability by the model with the U-correction suggests that most of these additional links correspond to true matches.  In particularly, essentially all of the mover matches identified by the Bayesian model discussed in Section~\ref{sec:false_match_rates} are assigned a non-trivial link probability only by the model with the U-correction.  

\section{Comparing linked and cross-sectional populations}\label{sec:linked-comp}

The lack of ground truth makes determining the representativness of the linked sample challenging to evaluate.  In particular, if false non-matches are concentrated among particular subgroups then bias may be introduced into the analysis of the linked data.  To examine this possibility we compare the marginal distributions of different subgroups observed in observed record pairs with those in the linked sample.  We report marginal distributions for gender and female marital status in Table~\ref{tab:marginal_gender_marital}. Table~\ref{tab:marginal_occupation_party} contains the marginal distributions for occupation and party membership. For the observed records counts and proportions are reported while for the linked sample we report the posterior mean estimated by our Bayesian model. 

The tables indicate that there is a high degree of similarity between the observed marginal distributions and those in the linked sample.  Table~\ref{tab:marginal_gender_marital} shows that in the linked sample men are over represented relative to women and married women are over represented relative to unmarried women. In both cases the findings are unsurprising as the over represented groups, men and married women, would generally be thought easier to link, relative to all women and unmarried women respectively. However, the discrepancy is a slight, with a maximum discrepancy between population shares is 0.03 in Table~\ref{tab:marginal_gender_marital}. The difference between observed and linked marginal distributions of white collar and blue collar workers is even smaller, just 0.01, as shown in the left panel of Table~\ref{tab:marginal_occupation_party}.  The right panel of Table~\ref{tab:marginal_occupation_party} shows the marginal distributions in party membership.  The differences between marginal distributions of party membership in the observed and linked samples are somewhat larger, up to 0.05, than in other subgroups.  However, we note that from 1932 to 1936 saw both an overall shift towards to Democratic party and a substantial increase in the number of registered voters.  By definition any new voters will be excluded from the linked sample since they are not present in the 1932 records.  If these new voters are heavily skewed towards the Democratic party, as it likely the case, then we would expected a representative linked sample to contain a higher share of Republican voters than the voter registrations overall. This is exactly what is observed with the observed and linked distributions of party membership for 1932 matching more closely than those for 1936.  Thus, while there are minor compositional differences between the observed and linked data in general the share of different subgroups display a surprisingly high level of agreememnt.

\begin{table}
\centering
\resizebox{0.99\textwidth}{!}{%
\begin{tabular}{lrrr}
  \hline
  \hline
Year & Record Count & All Men & All Women \\ 
  \hline
Observed 1932 & 259,162 & 0.53 & 0.47 \\ 
Observed 1936 & 288,087 & 0.51 & 0.49 \\ 
Linked 1932 & 137,640 & 0.54 & 0.46 \\ 
Linked 1936 & 137,640 & 0.54 & 0.46 \\ 
   \hline
   \hline
\end{tabular}\hspace{1em}%
\begin{tabular}{lrrr}
  \hline
  \hline
Year & Record Count & Married Women & Unmarried Women \\ 
  \hline
Observed 1932 & 122,596 & 0.82 & 0.18 \\ 
Observed 1936 & 142,034 & 0.82 & 0.18 \\ 
Linked 1932 & 63,553 & 0.83 & 0.17 \\ 
Linked 1936 & 63,922 & 0.84 & 0.16 \\ 
   \hline
   \hline
\end{tabular}
}
\caption{Marginal distribution of gender (left) and marital status for women (right) in observed and linked sample. For the linked sample posteriors means from our Bayesian model are reported. Both men and married women are slightly over represented in the linked sample indicting that these records are somewhat easier to link.}
\label{tab:marginal_gender_marital}
\end{table}

\begin{table}
\centering
\resizebox{0.99\textwidth}{!}{%
\begin{tabular}{lrrr}
  \hline
  \hline
Year & Record Count & Blue Collar Men & White Collar Men \\ 
  \hline
Observed 1932 & 63,619 & 0.92 & 0.08 \\ 
Observed 1936 & 66,921 & 0.92 & 0.08 \\ 
Linked 1932 & 35,474 & 0.91 & 0.09 \\ 
Linked 1936 & 33,501 & 0.91 & 0.09 \\ 
   \hline
   \hline
\end{tabular}\hspace{1em}%
\begin{tabular}{lrrr}
  \hline
  \hline
Year & Record Count & Democrat & Republican \\ 
  \hline
Observed 1932 & 232,106 & 0.33 & 0.67 \\ 
Observed 1936 & 266,465 & 0.53 & 0.47 \\ 
Linked 1932 & 125,890 & 0.30 & 0.70 \\ 
Linked 1936 & 128,508 & 0.48 & 0.52 \\ 
   \hline
   \hline
\end{tabular}
}
\caption{Marginal distribution of male occupation (left) and party (right) in observed and linked sample. For the linked sample posteriors means from our Bayesian model are reported. Both white collar men and republicans are slightly over represented in the linked sample.}
\label{tab:marginal_occupation_party}
\end{table}
\renewcommand{\thesection}{S.\arabic{section}}
\renewcommand{\thefigure}{S\arabic{figure}}
\renewcommand{\theequation}{S\arabic{equation}}
\setcounter{page}{0}
\setcounter{equation}{0}
\setcounter{table}{0}
\setcounter{figure}{0}
\setcounter{section}{0}

\title{Scaling Bayesian Probabilistic Record Linkage with Post-Hoc Blocking: An Application to the California Great Registers (Supplemental Material)}

\maketitle

\section{Post-Hoc Blocking Weights under One-to-One Constraints} \label{sec:assignment}

High-quality weights are important for efficient implementation of the post-hoc blocking algorithm.  In applications of PRL to historical data we often have relatively few fields available to perform matching, many or all of which are subject to error.  At the same time we know that the constituent files are at least approximately de-duplicated, so imposing a one-to-one matching constraint makes sense.  We also have limited or no labeled matching and non-matching record pairs with which to construct weights or validate results, suggesting the use of EM-estimated Fellegi-Sunter weights in post-hoc blocking.

However, we have observed that in this setting (one-to-one matching with a small number of noisy fields) the Fellegi-Sunter weights can be unreliable. We provide some evidence of this in Section~\ref{sec:assignment_italiam-census}. Similar observations have been made by \cite{tancredi2011hierarchical, sadinle2017bayesian}. In this section we propose a new method for estimating post-hoc blocking weights under one-to-one matching constraints by enforcing the constraints during estimation. 

\subsection{Penalized Maximum Likelihood Weight Estimates} \label{sec:assignment_penlike}

\cite{jaro1989advances}'s three-stage method for producing estimates of $C$ (summarized in Section~\ref{sec:background_fellegi-sunter:introduction}) involved three steps: Estimating the weights by maximum likelihood ignoring the one-to-one matching constraint, obtaining an optimal complete matching, and discarding matches with weights under a threshold to obtain a partial matching between the two files. Better estimates of the weights can be obtained by incorporating all three steps in a single-stage estimation procedure, simultaneously maximizing a joint likelihood in $C$, $m$, and $u$ while penalizing the total number of matches.

The penalized likelihood takes the following form, where the last term in \eqref{eq:base_penlik} is the penalty and the leading terms are the same complete data log likelihood corresponding to the standard two-component mixture model in \eqref{eq:fsmix}:
\begin{align}
    l(C, m, u; \Gamma) &= \sum_{ab}  [C_{ab}\log(m(\gamma_{ab})) + (1 - C_{ab})\log(u(\gamma_{ab}))]
    -\theta\sum_{ab} C_{ab} \label{eq:base_penlik}\\
    &= \sum_{ab} \log(u(\gamma_{ab})) + \sum_{ab}C_{ab}[\log(m(\gamma_{ab})) - \log(u(\gamma_{ab}))]
    -\theta\sum_{ab} C_{ab} \nonumber\\
    &= \sum_{ab} \log(u(\gamma_{ab})) + \sum_{ab}C_{ab}[w_{ab} - \theta]\label{eq:trans_penlik}
\end{align}    
%
The form of the penalized likelihood in \eqref{eq:trans_penlik} shows that $\theta$ plays a similar role to $T_\mu$ in the FS decision rule; only pairs with $w_{ab}>\theta$ can be linked without decreasing the log-likelihood.  This is also the unnormalized log posterior for $C$, $m$, and $u$ under the a prior for $C$ introduced in \cite{green2006bayesian}; the penalized likelihood estimate corresponds to a maximum a posteriori estimate under a particular Bayesian model.

Finding a local mode of \eqref{eq:base_penlik} is straightforward via alternating maximization steps, which are iterated until the change in \eqref{eq:base_penlik} is negligble:
\begin{enumerate}
    \item Maximize $C$, given values of $m$ and $u$. To maximize the penalized likelihood in $C$ we need to solve the following assignment problem:
\begin{equation} \label{eq:penlike}
\begin{aligned}
\max_{C}   &\sum_{a,b\in A\times B} C_{ab} \tilde w_{ab} \\
\subjectto\quad  & C_{ab} \in\{0, 1\}\\
& C_{ab} = 0\text{ if  }\tilde w_{ab}=0\\
&\sum_{b\in B} \hspace{0.5em} C_{ab} \leq 1 \hspace{2em} \forall a\in A \\
&\sum_{a\in A} \hspace{0.5em} C_{ab} \leq 1 \hspace{2em} \forall b\in B.
\end{aligned}
\end{equation}
where 
\begin{equation} \label{eq:pweight}
\tilde w_{ab} = \left\{\begin{array}{lr}
w_{ab} - \theta & w_{ab} \geq \theta\\
0 & w_{ab} < \theta
\end{array}\right.
\end{equation}
We discuss how to efficiently solve these thresholded assignment problems in Section~\ref{sec:assignment_maxweights_thresholding}.
\item Maximize $m$ and $u$ probabilities, given a value of $C$. These updates are available in closed form under the conditional independence model (Equation~\ref{eq:condindep}):

\begin{align}
m_{jh} &= 
\frac{
n_{mjh} + \sum_{ab} C_{ab}\ind{\gamma^j_{ab} = h}
}{
\sum_h n_{mjh} + \sum_{ab} C_{ab}
}\label{eq:Mest}\\
u_{jh} &= 
\frac{
n_{ujh} + \sum_{ab} (1-C_{ab})\ind{\gamma^j_{ab} = h}
}{
\sum_h n_{ujh} + \sum_{ab} (1-C_{ab})
}\label{eq:Uest}.
\end{align}
where the $n$'s are optional pseudocounts used to regularize the estimates. (These terms correspond to an additional penalty, omitted from \eqref{eq:base_penlik}-\eqref{eq:trans_penlik} for clarity.) We suggest their use in practice to avoid degenerate probabilities of zero or one. They are easy to calibrate as ``prior counts'' -- i.e., $n_{mjh}$ is the prior count of truly matching record pairs with level $h$ on comparison $j$, and the strength of regularization is determined by $\sum_h n_{mjh}$ (with larger values implying stronger regularization).
\end{enumerate}

Conceptually this optimization procedure is straightforward, but iteration to a global mode is not guaranteed.  In general a global mode in all the parameters need not exist -- for example, if a record in $A$ has two exact matches in $B$ then the penalized likelihood function will have at least two modes with the highest possible value. However, the values of the $m-$ and $u-$probabilities will be the same in both modes and these are the only objects of interest for defining weights. Of course it is also possible for an alternating maximization approach to get trapped in sub-optimal local modes.  Our experience running multiple starts from different initializations suggests that this not common -- that is, when we iterate to distinct local modes they tend to have similar values for the $m-$ and $u-$ probabilities.

\subsection{Maximal Weights for Post-Hoc Blocking} \label{sec:assignment_maxweights}

The estimated $m-$ and $u-$ probabilities obtained via penalized likelihood maximum estimation can depend strongly on the value of the penalty parameter $\theta$. In general higher values of $\theta$ correspond to lower numbers of matches, and one could potentially try to calibrate this parameter based on subject matter knowledge and prior expectations. However, rather than banking on our prior expectations we propose a more conservative approach: Rather than fixing a value of $\theta$ and obtaining weights for each pair, we vary $\theta$ over a range of values, obtain estimated weights for every value of $\theta$, and take the { maximum} observed weight for each record pair as the post-hoc blocking weight. This obviates the need to calibrate $\theta$ and assigns relatively high weight to any record pair that is a plausible match candidate for {\em some} value of $\theta$. 

To define the sequence of values we suggest starting with $\theta = 0$ and then selecting successively larger penalty values.  The actual sequence of penalty values can be chosen via a variety of different rules.  A useful rule of thumb is that the next penalty in the sequence should be larger than the smallest weight in the previous solution, to ensure a change in the solution to the assignment problem.  Specifying a minimum gap size between successive values of $\theta$ provides further control over computation time.  

A naive implementation of maximal weight estimation is computationally intensive -- standard algorithms for solving the assignment problems (such as the Hungarian algorithm, \cite{kuhn1955hungarian}) have worst-case complexity that is cubic in the larger of the two file sizes.  Each step of the penalized likelihood maximization involves solving multiple assignment problems, and this must be repeated for each distinct value of $\theta$. However, there are three features of our assignment problems that make them dramatically easier to solve: The weight matrices involved are usually extremely sparse, exact solutions are often not necessary, and assignments from previous iterations can be used to effectively initialize the next iteration.

\subsubsection{Solving Sparse Thresholded Assignment Problems} \label{sec:assignment_maxweights_thresholding}

Given a set of estimated weights, \cite{jaro1989advances} suggested solving \eqref{eq:jaroLSAP_2} by constructing a canonical linear sum assignment problem, which assumes that each record in $A$ will be matched to some record in $B$. If $n_A<n_B$ \cite{jaro1989advances} does this by constructing an $n_B\times n_B$ augmented square matrix of weights $\check{w}$ and an augmented assignment matrix $C$ and solving the following canonical {\em linear sum assignment problem} (LSAP):
 \begin{equation} \label{eq:jaroLSAP_2}
 \begin{aligned}
  \max_{C}   &\sum_{a=1}^{n_B}\sum_{b=1}^{n_B}  C_{ab}\check{w}_{ab}& \\  
 \subjectto\quad  & C_{ab} \in\{0, 1\}\\
 &\sum_{b=1}^{n_B} \hspace{0.5em} C_{ab} = 1 \hspace{2em} \forall a\in A \\
 &\sum_{a=1}^{n_A} \hspace{0.5em} C_{ab} = 1 \hspace{2em} \forall b\in B,
 \end{aligned}
 \end{equation}
 where $\check{w}_{ab} = \hat{w}_{ab}$ (the estimated weight) if $a\leq n_A$ and is otherwise set to the smallest observed weight or another extreme negative value. In a final step  any matches with weights under a threshold are dropped, which necessarily removes any matches that correspond to augmented entries in $C$.

\begin{figure*}
\centering
\captionsetup[subfigure]{justification=centering}
\begin{subfigure}{.2\textwidth}
\centering
\resizebox{0.95\textwidth}{!}{%
\begin{tikzpicture}

\node[] at (-0.4,1.5) {$\mathbf{a}_1$};
\node[] at (-0.4,0.5) {$\mathbf{a}_2$};
\node[] at (0.5,2.4) {$\mathbf{b}_1$};
\node[] at (1.5,2.4) {$\mathbf{b}_2$};
\node[] at (2.4,0.0) {};

\node[] at (0.5,0.5) {$\mathbf{2}$};
\node[] at (1.5,0.5) {$\mathbf{-3}$};
\node[] at (0.5,1.5) {$\mathbf{5}$};
\node[] at (1.5,1.5) {$\mathbf{2}$};

\draw[step=1cm,black,thick] (0,0) grid (2,2);
\end{tikzpicture}
}%
\caption{}
\label{penlsap:costs}
\end{subfigure}\hspace{.05\linewidth}%
\begin{subfigure}{.2\textwidth}
\centering
\resizebox{0.95\textwidth}{!}{%
\begin{tikzpicture}

\filldraw[fill=green!60!black] (0,0) rectangle (1,1);
\filldraw[fill=green!60!black] (1,1) rectangle (2,2);

\node[] at (-0.4,1.5) {$\mathbf{a}_1$};
\node[] at (-0.4,0.5) {$\mathbf{a}_2$};
\node[] at (0.5,2.4) {$\mathbf{b}_1$};
\node[] at (1.5,2.4) {$\mathbf{b}_2$};
\node[] at (2.4,0.0) {};

\node[] at (0.5,0.5) {$\mathbf{2}$};
\node[] at (1.5,0.5) {$\mathbf{-3}$};
\node[] at (0.5,1.5) {$\mathbf{5}$};
\node[] at (1.5,1.5) {$\mathbf{2}$};
\draw[step=1cm,black,thick] (0,0) grid (2,2);

\end{tikzpicture}
}%
\caption{}
\label{penlsap:sol1}
\end{subfigure}\hspace{.05\linewidth}%
\begin{subfigure}{.2\textwidth}
\centering
\resizebox{0.95\textwidth}{!}{%
\begin{tikzpicture}

\node[] at (-0.4,1.5) {$\mathbf{a}_1$};
\node[] at (-0.4,0.5) {$\mathbf{a}_2$};
\node[] at (0.5,2.4) {$\mathbf{b}_1$};
\node[] at (1.5,2.4) {$\mathbf{b}_2$};
\node[] at (2.4,0.0) {};

\node[] at (0.5,0.5) {$\mathbf{2}$};
\node[red] at (1.5,0.5) {$\mathbf{0}$};
\node[] at (0.5,1.5) {$\mathbf{5}$};
\node[] at (1.5,1.5) {$\mathbf{2}$};
\draw[step=1cm,black,thick] (0,0) grid (2,2);

\end{tikzpicture}
}%
\caption{}
\label{penlsap:pencosts}
\end{subfigure}\hspace{.05\linewidth}%
\begin{subfigure}{.2\textwidth}
\centering
\resizebox{0.95\textwidth}{!}{%
\begin{tikzpicture}

\filldraw[fill=blue!60!white] (0,1) rectangle (1,2);

\node[] at (-0.4,1.5) {$\mathbf{a}_1$};
\node[] at (-0.4,0.5) {$\mathbf{a}_2$};
\node[] at (0.5,2.4) {$\mathbf{b}_1$};
\node[] at (1.5,2.4) {$\mathbf{b}_2$};
\node[] at (2.4,0.0) {};

\node[] at (0.5,0.5) {$\mathbf{2}$};
\node[] at (1.5,0.5) {$\mathbf{0}$};
\node[] at (0.5,1.5) {$\mathbf{5}$};
\node[] at (1.5,1.5) {$\mathbf{2}$};
\draw[step=1cm,black,thick] (0,0) grid (2,2);

\end{tikzpicture}
}%
\caption{}
\label{penlsap:sol2}
\end{subfigure}
\caption{Simple assignment problem with and without threshold (\subref{penlsap:costs}) shows an example of estimated weights. (\subref{penlsap:sol1}) Highlights the assignment that maximizes the assigned weights for the costs given in (\subref{penlsap:costs}).  (\subref{penlsap:pencosts}) Adjusts the costs shown in (\subref{penlsap:costs}) by thresholding at 0.  (\subref{penlsap:sol2}) the maximal assignment solution if the thresholded costs are used and zero cost assignments are then deleted.  We note that the resulting assignment has a higher weight than the one given in (\subref{penlsap:sol1}).}
\label{fig:penlsap}
\end{figure*}
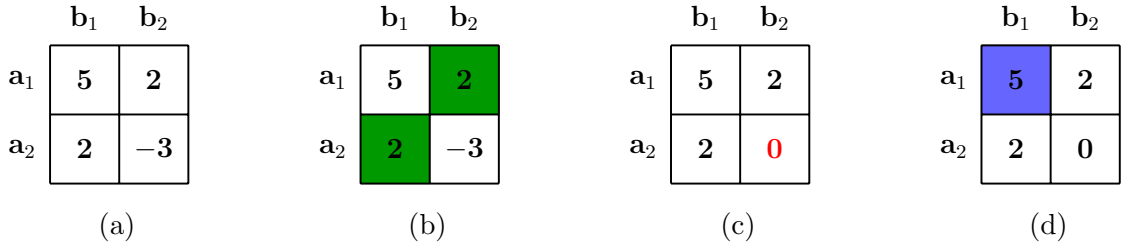

Unfortunately, in general this procedure will not lead to the estimate of $C$ with the highest total weight assigned to the matched pairs. Figure~\ref{fig:penlsap} provides a simple counterexample. Figure \ref{penlsap:costs} shows an example of estimated weights. Since the LSAP above makes a complete assignment, there are two feasible values of $C$ that could be returned: Either $a_1$ matches $b_1$ and $a_2$ matches $b_2$, or $a_1$ matches $b_2$ and $a_2$ matches $b_1$.  The latter matching (Figure \ref{penlsap:sol1}) provides the solution to the assignment problem above, because of the relatively large negative weight on the pair $(a_2, b_2)$. But inspecting the weight matrix shows that the best {\em overall} matching -- accounting for the subsequent thresholding -- is obtained by linking $a_1$ and $b_1$, leaving $a_2$ and $b_2$ unmatched.

The solution to this problem is to incorporate the threshold into the maximization problem by setting any weights below the threshold to zero (and adding a constant to make the remaining weights positive if necessary). In the final step, any entry of $C$ with a corresponding zero weight is dropped. Figure \ref{penlsap:pencosts} shows the thresholded weight matrix, which leads to the correct solution (Figure \ref{penlsap:sol2}). This is how we construct the weight matrices during penalized likelihood estimation; see \eqref{eq:pweight}.

Adopting the formulation in \eqref{eq:penlike} has the added benefit of making the assignment problem easier to solve.  While relatively efficient algorithms exist for solving dense LSAPs, (e.g. the Hungarian algorithm \citep{kuhn1955hungarian}), they have a worst case complexity of $O(n^3)$ where $n = \max(n_A, n_B)$ \citep{jonker1986improving, lawler1976combinatorial}.  However, after thresholding our weight matrix will be very sparse. Indeed, depending on the degree of overlap between the two files there may be entire rows and columns of zeros -- effectively reducing $n$ and yielding an easier optimization problem.

Even greater benefits can be realized by partitioning the sparse weight matrix to derive many small optimization problems to be solved in parallel. Similar to our procedure for obtaining post-hoc blocks, we can employ graph clustering to separate the records into blocks (defined by the connected components of the weighted graph defined by the thresholded post-hoc blocking weights) such that links are only possible within and not across blocks. This allows us to decompose the full assignment problem into a set of smaller problems that can be solved in parallel, as summarized in Algorithm~\ref{alg:componentassignment}.

\begin{algorithm}
\caption{Connected-Component Based Assignment Problem}
\label{alg:componentassignment}
\begin{algorithmic}
\INPUT{Thresholded weight matrix $\widetilde W$ (from Eq~\eqref{eq:pweight})}
\OUTPUT{Estimate $\widehat C$ partial assignment with highest total weight}
  \begin{enumerate}
    \item Find the connected components of the bipartite graph $G$ which has edges between nodes $a$ and $b$ where $\tilde w_{ab}>0$.
    \item Solve the assignment problem for each component separately.
    \item Merge assignment solutions.
  \end{enumerate}
%
\end{algorithmic}
\end{algorithm}

Finding the connected components of a bipartite graph has computational complexity of $O(|E| + n_A + n_B)$ (linear time with respect to the number of edges in the graph, i.e. the number of nonzero weights after threshholding) \citep{tarjan1972depth}. This is loosely bounded by $n_An_B$ above.  After partitioning the graph, computational demands are driven primarily by solving the LSAP corresponding to the largest connected component. Since the computational complexity of this step is at worst $O(k^3)$, with $k$ being the maximum number of records from either file appearing in the component, we can obtain dramatic reductions in computational complexity by partitioning the original problem.

Practical performance is often much better than these worst-case complexity results might suggest. Many of the sub-matrices of $\tilde W$ corresponding to connected components will remain sparse. In computational studies many algorithms for solving LSAPs show substantially faster results on sparse problems \citep{carpaneto1983algorithm,jonker1987shortest,orlin1993quickmatch,hong2016solving}.  In fact previous work suggests, but does not prove, that it may be possible to solve sparse assignment problems in near linear time with respect to the number of edges \citep{orlin1993quickmatch}. The only case of proven complexity improvements that we are aware of is for auction algorithms \citep{bertsekas1988dual,bertsekas1989parallel}.

We adopt the auction algorithm for solving sparse LSAPs in all of our algorithms.  In addition to its performance guarantees, the auction algorithm allows us to use previous solutions to specify initial values for new problems. This is useful in the iterative maximization problems in both the penalized maximum likelihood and the maximal weight procedure.  Further, we have the option to stop the auction algorithm early to save on computation time. This is useful in penalized likelhood estimation, where we need not necessarily find the optimal assignment in order to improve the objective function at each step.  For a more complete overview of auction algorithms see \cite{bertsekas1989parallel, bertsekas1998network, bertsekas1992auction}.  
\textbf{}

\subsection{Illustrations of Post-Hoc Blocking and Restricted MCMC: Italian Census Data \citep{tancredi2011hierarchical}}\label{sec:assignment_italiam-census}

We consider a small scale example from the existing literature to illustrate the performance of post-hoc blocking with maximal weights.  The data in this example come from a small geographic area; there are 34 records from the census (file A) and 45 records from the post-enumeration survey (file B). The goal is to identify the number of overlapping records to obtain an estimate of the number of people missed by the census count using capture-recapture methods.  This small scale example allows us to compare the results of estimation performed employing post-hoc blocking with the results from considering the full set of record pairs.

Each record includes three categorical variables: the first two consonants of the family name (339 categories), sex (2 categories), and education level (17 categories). We generate comparison vectors as binary indicators of an exact match between each field. The prior over the linkage structure is set to a Beta-bipartite distribution with $\alpha=1.0$ and $\beta = 1.0$, which is uniform over the expected proportion of matches \citep{fortini2001bayesian,fortini2002modelling,larsen2005advances,larsen2010record,sadinle2017bayesian}. We assume a conditional independence model for $m-$ and $u-$probabilities as in \eqref{eq:condindep}. Each vector of conditional probabilities is assigned a Dirichlet prior distribution. We assume that $m_j\sim \mathrm{Dir}(1.9, 1.1)$ and $u_j\sim \mathrm{Dir}(1.1, 1.9)$ for $j = 1, 2, 3$ independently.  These priors where chosen to contain modes near 0.9 and 0.1 respectively, with a reasonable degree of dispersion.  

\begin{table}
  \centering
  \begin{tabular}{rrrrrr}
    \hline
    \hline
    Last & Sex & Edu & Count & EM Weight & Maximum Weight\\
    \hline
    1 & 1 & 1 &  25 &  5.27 &  6.21 \\
    1 &   0 &   1 &   8 & 3.77 & 3.04 \\ 
    \rowcolor{Gray}
    1 &   1 &   0 &  13 & -0.94 & 2.23 \\ 
    \rowcolor{Gray}
    0 &   1 &   1 & 126 & 2.68 & 0.24 \\ 
    1 &   0 &   0 &  21 & -2.45 & -1.03 \\ 
    \rowcolor{Gray}
    0 &   0 &   1 &  78 & 1.18 & -3.14 \\ 
    0 &   1 &   0 & 601 & -3.53 & -3.81 \\ 
    0 &   0 &   0 & 658 & -5.04 & -7.19 \\ 
    \hline
    \hline
  \end{tabular}
  \caption{Maximum weights used for post-hoc blocking, and EM weights for comparison}
  \label{tab:weights}
\end{table}

We estimate post-hoc blocking weights using the maximal weight procedure in Section~\ref{sec:assignment_maxweights}.  The resulting weights for each possible comparison vector are shown in Table~\ref{tab:weights}, along with EM weights for comparison. Notable discrepancies are in gray. The EM weights consider a record pair agreeing on sex and education alone to be a more probable match than a record pair agreeing on last name and sex, assigning such record pairs weights of 2.68 and -0.94 respectively. While we have no ground truth here, this seems unlikely. More striking is the EM weight assigned to record pairs that agree solely on education -- its value of 1.18 under this model would suggest that these such a pair is more likely than not a true match. Overall the maximum weights seem to provide a more reasonable rank ordering of the comparison vectors.

Given the small size of the problem we select only a single post-hoc blocking threshold $w_0$ to implement the restricted MCMC.  In our post-hoc blocking procedure we limit the size of the largest post-hoc block to fewer than 100 record pairs. The resulting post-hoc blocks contain only 94 of the 1530 possible record pairs.  These are spread across 21 separate post-hoc blocks.  Of the 21 post-hoc blocks, 14 contain only a single record pair, 4 contain 2 record pairs, the remaining three contain 4, 8, and 60 record pairs respectively.  

\begin{figure}
  \centering
  \begin{subfigure}{.539\linewidth}
    \centering
    \includegraphics[scale=0.98]{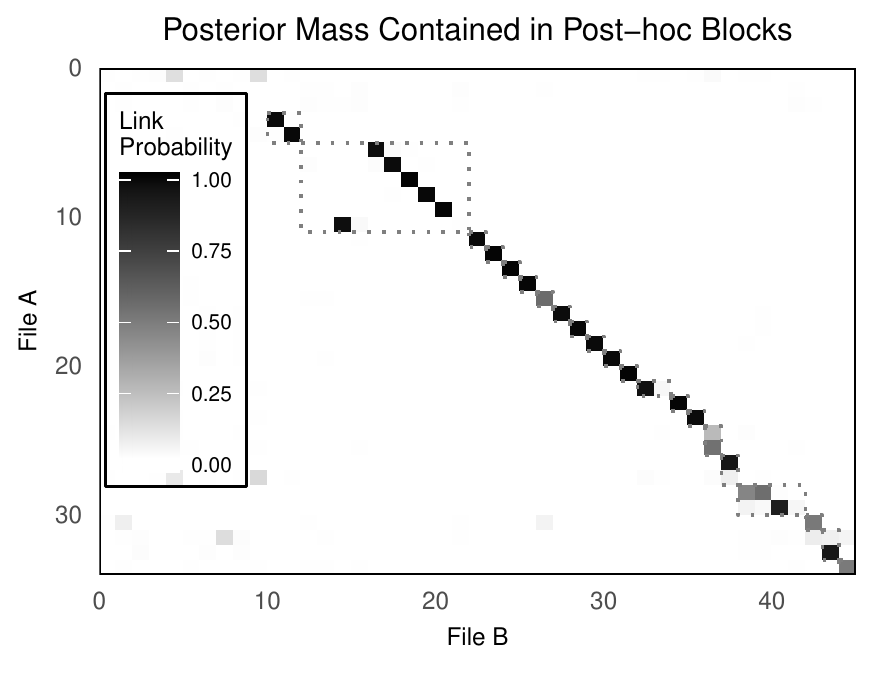}
    \caption{}
    \label{tl_results:posthoc_blocks}
  \end{subfigure}%
  \begin{subfigure}{.459\linewidth}
    \centering
    \includegraphics[scale=0.98]{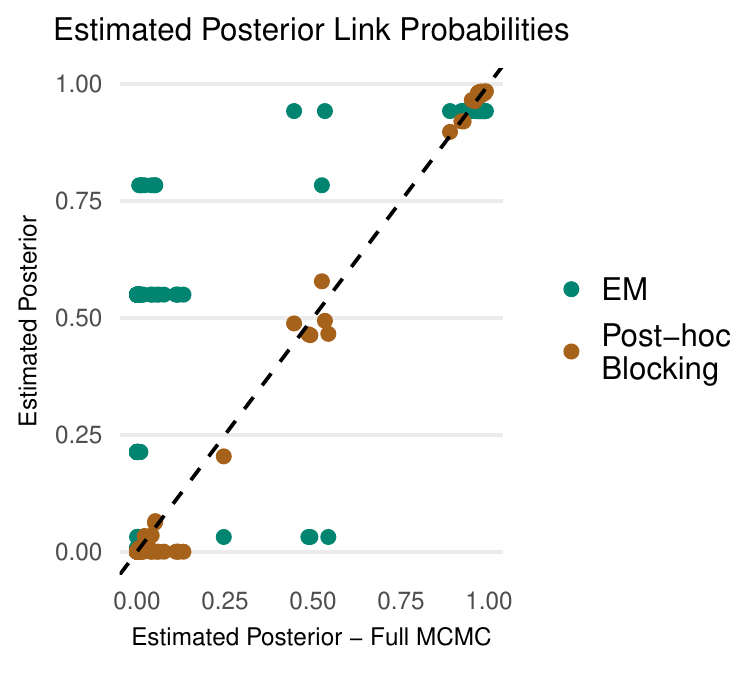}
    \caption{}
    \label{tl_results:fullvblocks_prob}
  \end{subfigure}
      \caption{(a) Post-hoc blocks overlayed on posterior link probabilities estimated via MCMC using all record pairs (b) Posterior probabilities from EM and restricted MCMC versus posterior match probability considering all record pairs.}
      \label{fig:tl_results}
\end{figure}

We then run both a MCMC algorithm containing all 1530 record pairs and our restricted MCMC under identical model specifications. Results from both models are displayed in  Figure~\ref{fig:tl_results}\subref{tl_results:posthoc_blocks}, with the post-hoc blocks overlayed.  Nearly all of the posterior link density is contained within the post-hoc blocks, but a few pairs with modest posterior probability are omitted from the post-hoc blocks. (Lowering the threshold to capture these would have resulted in a single large block.)

In Figure~\ref{fig:tl_results}\subref{tl_results:fullvblocks_prob} we compare the posterior match probability estimated by the full MCMC, our post-hoc blocking restricted MCMC, and posterior probability estimates as computed from the EM output.  The full and restricted MCMC probabilities are quite similar, except the small cluster of points on the x-axis near the origin. These are points that had modest posterior probability -- less than 0.12 -- in the full MCMC but were excluded from the post-hoc blocks and assigned zero probability in the approximate posterior. Even in this small example we obtain a significant improvement in runtimes: Using identical implementations posterior sampling takes 3.4 seconds for the full MCMC algorithm versus 0.32 seconds when employing post-hoc blocks.  The order of magnitude reduction in runtime is almost certainly an understatement if we also consider the mixing time of the two chains -- the restricted chain targets its moves carefully and tends to mix much faster.  

The EM fit provides estimates of posterior probabilities, albeit posterior probabilities that do not respect one-to-one matching constraints. These estimates do not align well with the MCMC output. This is in part due to the problematic weight estimates in Table~\ref{tab:weights}. But the failure to account for one-to-one matching seems to play a larger role -- in general we would expect omitting the constraint to lead the posterior probability estimates to be too high, which is what we see here -- nearly all the EM posterior probabilities exceed the Bayesian estimates.

\section{Extended Comparison of fastLink and the Bayesian Model}\label{sec:fastlink-compariosns}

From the results presented in Section~\ref{sec:false_match_rates} it is clear that the estimates produced by the Bayesian model (Section~\ref{sec:alameda_bayesian}) and fastLink (Section~\ref{sec:alameda_fastlink}) differ substantially. To help understand why these differences arise, here we fit a version of the Bayesian model that is closer to fastLink in its specification. We find that the performance gap between fastLink and the Bayesian model persists; in Section~\ref{sec:alameda_method_diff} we discuss the remaining differences and their potential impacts on our results.

\subsection{A Comparable Bayesian Model}\label{sec:comparable-model}
The Bayesian model and fastLink differ not only in the estimation procedure used, but also in the comparison vectors computed (due to constraints in the fastLink procedure) and even in the specific set of record pairs considered (due to differences in the blocking and indexing schemes).  Thus, it is not immediately obvious from the examination of the presented faslse match rates which of these differences is the driver for the lower false match rates produced by the Bayesian model.  We therefore estimate the linkage structure of the Alameda county voter files using a second Bayesian model, our ``Comparable'' model, on the set of record pairs and comparison vectors used by the fastLink estimation procedure.  As described in Section~\ref{sec:alameda_fastlink} we block on first name using internal fastLink functions and compute comparison vectors using the  ``exact'' match, partial match, and non-match bins defined by default values from fastLink.  We do however share the model parameter estimates across blocks as in the Bayesian model, whereas fastLink estimates model parameters separately within each block.

As with out Bayesian model we construct post-hoc blocks via estimated maximal weights using a sequence of penalized likelihood estimators setting $N_c$ to 250,000 and initializing the post-hoc blocking procedure with a $w_{min}$ value of zero.  This produces a set of 86,000 distinct post-hoc blocks containing 21.7 million record pair, somewhat fewer post-hoc blocks but substantially more record pairs than the post-hoc blocks than our Bayesian model.  The cause for this appears to be the smaller number of bins used in the fastLink comparison vectors making the maximal weight matrix ``flatter'', taking fewer distinct values, and therefore harder to separate into small post-hoc blocks.

For the restricted MCMC algorithm the prior over the $m$-parameters is set to $m_j\sim$Dir(10,5,1) for first name, middle name, surname, occupation, street number, and street name, the fields for which partial matches are computed.  For female and street type, where only exact matches and non-matches are computed, we set the prior to $m_j\sim$Dir(10,1). For the $u$-parameters flat priors of $u_j\sim$Dir(1,1,1) and $u_j\sim$Dir(1,1) are used for partial matching and exact matching fields respectively. As with our Bayesian model we employ a Beta-bipartite prior with $\alpha = 1.0$ and $\beta=1.0$ over the link structure and run the MCMC algorithm for 25,000 steps, discarding the first 2,500 steps as burn-in.  We run the estimation procedure with the U-Correction which, as discussed in Section~\ref{sec:alameda_bayesian_estimate}, is essential to correctly evaluating the likelihood.  

We begin our comparisons of the model results with an examination of the differences in the indexing schemes. The blocking used for fastLink and the Comparable model consider a total of 1.1 billion record pairs, compared to the 850 million record pairs included in the indexing scheme developed for the Bayesian model.  While there is a substantial degree of overlap, a total of 630 million pairs appear in both indexing schemes, the overlap among record pairs with a high posterior link probability is more substantial. Of the 132,331 record pairs for which the Bayesian model estimates a posterior link probability greater than 0.5, 119,625 (90.3\%) are included in the fastLink blocking scheme. The Bayesian model indexing scheme
includes an even greater fraction of the links estimated by the Comparable model, containing 116,345 (99.9\%) of the 117,477 record pairs assigned a posterior match probability greater than 0.5 by this model. The overlap with the fastLink results is less but still substantial, with the Bayesian model indexing scheme containing only 65,035 (82.1\%) of the 79,228 record pairs assigned a posterior link probability greater than 0.9 by fastLink.

\begin{figure}
  \centering
  \includegraphics[scale=1.0]{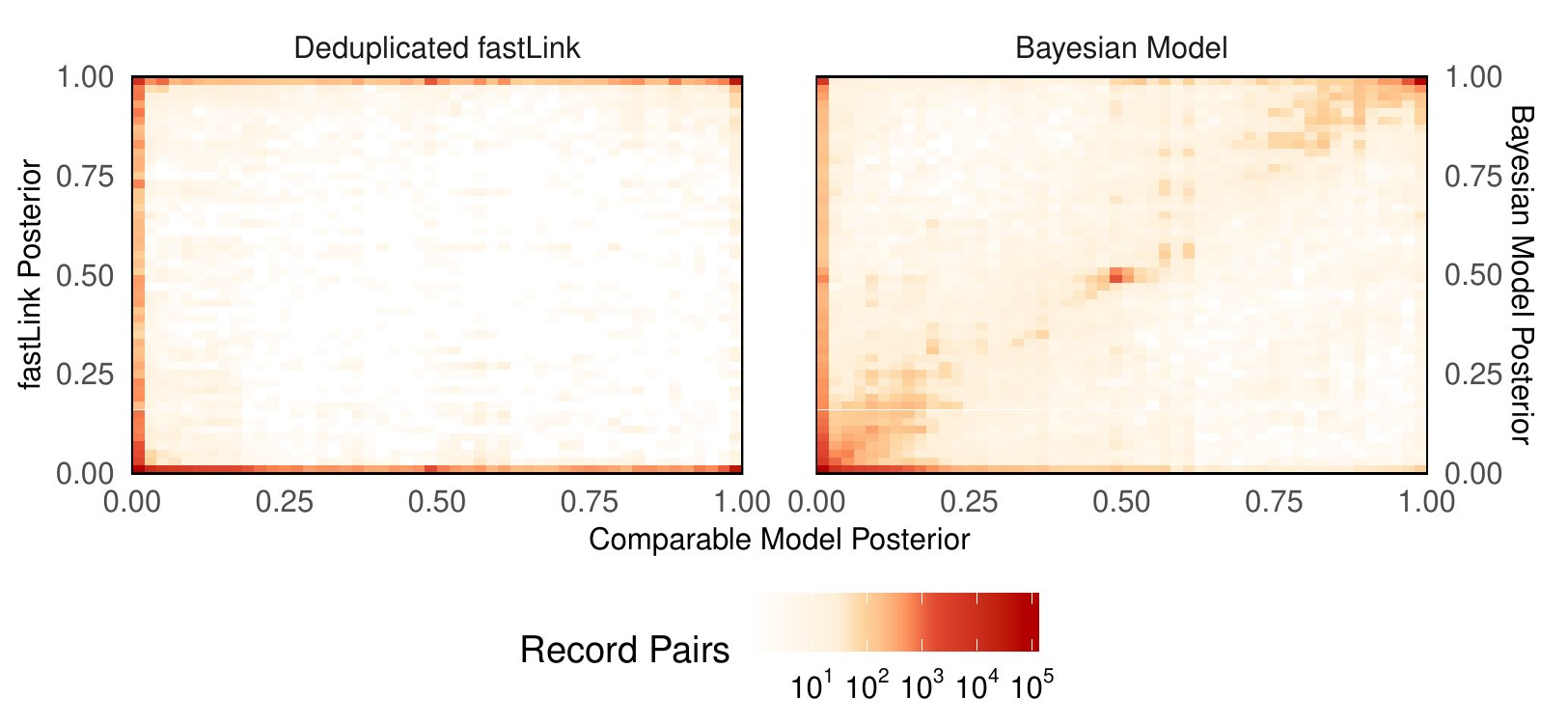}
  \caption{Heat map of estimated posterior link probabilities Comparable model vs fastLink (left) and Comparable model vs. Bayesian model (right).}
  \label{fig:density_comparable}
\end{figure}

We next examine the posterior link probabilities estimated by the models at the record pair level.  After first excluding all record pairs for which all three models (Bayesian, Comparable, and fastLink) estimate less a posterior link probability of less than .0001 we plot a heat map showing counts on a log scale of the posterior link probabilities estimated by the different models in Figure~\ref{fig:density_comparable}.  We compare the probabilities estimated by fastLink with the Comparable model in the left panel and the Comparable model with the Bayesian model in the right panel.  Record pairs for which the models estimate similar posterior link probabilities appear near the diagonal while those where the models estimated link probabilities which differ substantially appear closer to the top left and bottom right corner.

It is clear that between the Comparable model and fastLink there is essential zero agreement on posterior link probability, outside of the near certain non-links (bottom left) and near certain links (top right) record pairs.  In contrast, the Bayesian model and the Comparable model plot shows substantial mass near the diagonal, indicating broad agreement on the posterior link probabilities.  This occurs despite the models relying on different comparison vectors.  There does exist a set of several thousand record pairs for which the Comparable model estimates zero link probability which the Bayesian model classifies as near certain links (shown in the top left corner of the right panel).  However, upon further examination we determined that this disagreement is the result of differences in indexing scheme already described rather than different estimates for record pairs included in both models.  Overall it is clear that there is substantially more agreement between the Bayesian and Comparable models, both estimated using our post-hoc blocking procedure, suggesting that the choice of estimation procedure is significantly more important in determining model performance than specifics of the comparison vectors or blocking scheme.

\subsubsection{False Match Rates}\label{sec:comparable-model_fmr}

As a final comparison between the modeling approaches we expend the false match rate analysis from Section~\ref{sec:false_match_rates} to cover the Comparable model.  Links identified by the Comparable model appear in all strata, Bayesian only, fastLink only, and intersection as well as a small number, 3,626, which are linked by neither fastLink nor the Bayesian model and thus appear in none of the previous strata.  We therefore label 200 additional records pairs, 100 mover and 100 non-mover matches, identified (assigned a posterior link probability greater than 0.5) by the Comparable model which are classified as non-matches by both the Bayesian model (using a threshold of 0.5) and fastLink (using a threshold of 0.9) in order to estimate the false match rate achieved by the Comparable model.  We reproduce Table~\ref{tab:labeling} with the additional labels added in Table~\ref{tab:labeling_comparable}.  

\begin{figure}[h!]
  \centering
  \includegraphics[scale=1.0]{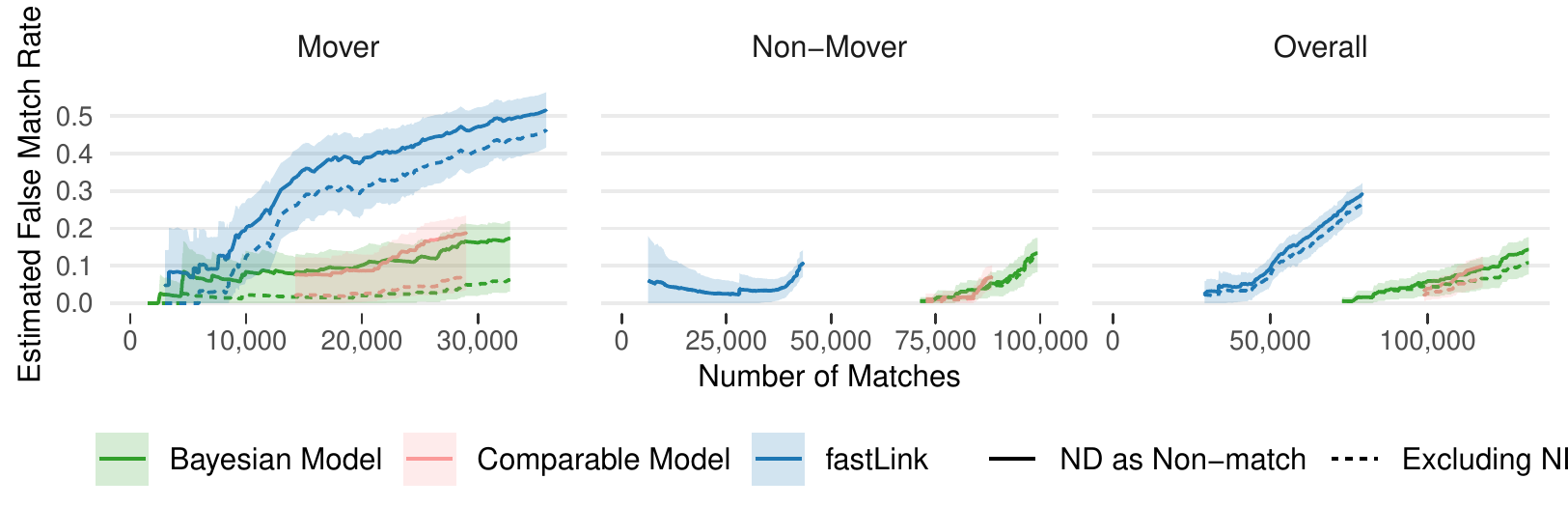}
  \caption{Estimated false match rates for mover matches (left), non-mover matches (center), and all matches (right) for deduplicated fastLink (blue) and Bayesian model (green). False match rate relative to hand-coding base estimates excluding ND record pairs (dashed) and counting ND record pairs as non-matches (solid).}
  \label{fig:fmr_all_comparable}
\end{figure}

We then repeat the analysis in Section~\ref{sec:false_match_rates} including the Comparable model updating Table~\ref{tab:model_fmr_estimates} and Figure~\ref{fig:fmr}
in Table~\ref{tab:model_fmr_estimates_comparable} and Figure~\ref{fig:fmr_all_comparable} respectively.  Figure~\ref{fig:fmr_all_comparable} in particular shows that while the Bayesian model identifies some additional matches, largely due to the blocking scheme, the performance of the two models is extremely similar with substantial overlap between the estimated false match rates.  These results should be unsurprising given the agreement observed in Figure~\ref{fig:density_comparable} but do provide yet more evidence that, at least for this problem, the estimation procedure matters far more to the modeling than do the specific comparison vectors or blocking scheme.

\begin{table}
\centering
\resizebox{0.99\textwidth}{!}{%
\begin{tabular}{l|rrrrr|rrrrr|rrrrr}
\hline
\hline
\multicolumn{1}{c}{} & \multicolumn{5}{c}{Mover} & \multicolumn{5}{c}{Non-Mover} & \multicolumn{5}{c}{Overall} \\
Stratum & FM & TM  & ND & Labeled & Total Matches & FM & TM  & ND & Labeled & Total Matches & FM  & TM  & ND & Labeled & Total Matches \\
\hline
Intersection &   2 &  88 &  10 & 100 & 14,276 &   4 &  96 &   0 & 100 & 38,968 &   6 & 184 &  10 & 200 & 53,244 \\ 
Bayesian Only &  12 & 118 &  20 & 150 & 18,525 &  26 & 121 &   3 & 150 & 60,562 &  38 & 239 &  23 & 300 & 79,087 \\ 
fastLink Only & 102 &  33 &  15 & 150 & 21,636 & 105 &  44 &   1 & 150 & 4,348 & 207 &  77 &  16 & 300 & 25,984 \\ 
Comparable Only &  32 &  53 &  15 & 100 & 2,213 &  68 &  28 &   4 & 100 & 1,413 & 100 &  81 &  19 & 200 & 3,626 \\ 
\hline
\hline
\end{tabular}
}
\caption{Hand-coding results from mover (left) and non-mover (center) matches and overall (right).  Each matched record pair was labeled as either a false match (FM), a true matches (TM) or no determination (ND), when insufficient information was available.}
\label{tab:labeling_comparable}
\end{table}

\begin{table}
\centering
\resizebox{0.95\textwidth}{!}{%
\begin{tabular}{l|ccc|ccc}
\hline
\hline
\multicolumn{1}{c}{} & \multicolumn{3}{c}{ND Excluded} & \multicolumn{3}{c}{ND as Non-match} \\
 & Mover & Non-Mover & Overall & Mover & Non-Mover & Overall \\
\hline
Bayesian Model & 0.062 (0.031, 0.093) & 0.123 (0.083, 0.164) & 0.108 (0.077, 0.139) & 0.173 (0.126, 0.219) & 0.133 (0.092, 0.175) & 0.143 (0.110, 0.176) \\ 
  fastLink & 0.464 (0.419, 0.509) & 0.107 (0.071, 0.142) & 0.269 (0.240, 0.297) & 0.518 (0.470, 0.565) & 0.107 (0.072, 0.142) & 0.293 (0.264, 0.322) \\ 
  Comparable Model & 0.072 (0.038, 0.107) & 0.066 (0.036, 0.095) & 0.067 (0.044, 0.091) & 0.188 (0.142, 0.234) & 0.070 (0.040, 0.101) & 0.099 (0.074, 0.125) \\ 
\hline
\hline
\end{tabular}
}
\caption{Estimated false match rates with 95\% confidence intervals excluding ND record pairs (left) and counting ND record pairs as false matches (right) by model.}
\label{tab:model_fmr_estimates_comparable}
\end{table}

\subsection{Examining other differences between the methods}\label{sec:alameda_method_diff}

Since choices about string similarities and blocking procedures do not explain the gap in results between the Bayesian model and fastLink; we discuss some important remaining differences between the methods below, and how they might impact performance.

\begin{figure}
  \centering
  \includegraphics[scale=1.0]{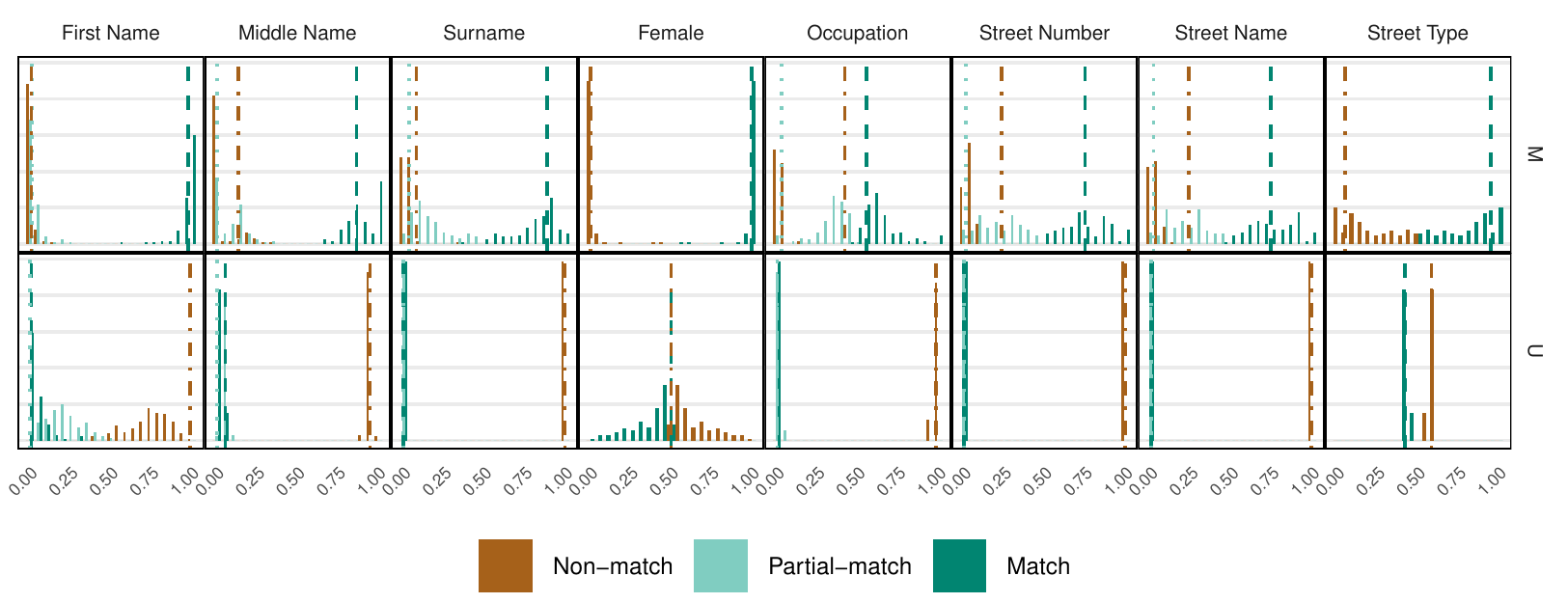}
  \caption{Distribution of fastLink parameters across blocks. Vertical lines show posterior means of parameters for the Comparable model.}
  \label{fig:parameters_fastlink}
\end{figure}

\noindent\textbf{Common versus distinct parameters and the $U-$correction.} fastLink estimates a distinct Fellegi-Sunter model (Eq.~\eqref{eq:fsmix}) via regularized EM within each block, yielding 123 separate estimates for each model parameter. Since fastLink does not make a $U-$correction this model makes some sense, as we would expect the at least some of the $u-$probabilities  -- which roughly measure the probability that two randomly selected records within a block will agree or partially agree by chance -- to vary across blocks. (It is less clear why the $m-$ probabilities, which capture measurement/recording or reporting error, should vary by blocks defined by the first name field.) Figure~\ref{fig:parameters_fastlink} shows histograms of these parameter estimates along with the posterior means from the comparable Bayesian model (labeled ``Comparable''). We see the expected variability of fastLink's first name $u-$parameters across blocks. We see similar variability in the $u-$parameter for female, since the distribution of gender varies across fastLink's blocks. There is marked variability in the fastLink estimated $m-$probabilities across blocks, which is more difficult to explain by anything other than sampling variability (even though many of the blocks are large, the proportion of matching pairs is small and fastLink does not borrow information about the $m-$probabilities across blocks).

\noindent\textbf{Constraints on the parameter estimates.} fastLink  imposes the following constraints on the parameter estimates: $m_{match} \geq m_{partial-match} \geq m_{non-match}$ and $u_{match} \leq u_{partial-match} \leq u_{non-match}$ \citep{enamorado2018using}. It does not appear that this assumption is reasonable in our application. Consider the case where random agreement on a field is relatively common but transcription errors are either uncommon or, more plausibly, result in the a comparison falling into the lowest similarity bucket.  In this scenario partial agreements may be observed less frequently than either matching comparisons among non-matched record pairs (due to random agreement) or non-matching comparisons among matched record pairs (due to parsing or transcription errors).  This would violate the constraints imposed by fastLink and result in parameter estimates on the boundary.

Concretely we see these effects in the Bayesian $m-$probability estimates for occupation and the address fields. Conditional on the record pair being a true match, it is most likely that the street name, number and occupation agree, but the next most likely outcome is {\em disagreement}, not partial agreement. This makes sense, as occupations are often inconsistently recorded in a fashion leading to low similarity scores (a ``stevedore'' in 1932 might report his occupation as a ``dockworker'' in 1936) and addresses are subject to seemingly random failures in the OCR and parsing. We might correct some of these effects with better preprocessing (e.g. more intensive standardization of occupations), but both addresses and occupations are subject to change over time. If these changes occur at a higher rate than minor typographical or OCR errors leading to partial agreement then the constraint would still be violated.

Finally, note that as long as partial matches are \textit{relatively} more common among truly matching record pairs than non-matching pairs then an observed partial match will still, other features held constant, indicate that the record pair is more likely to be a true match.  Thus, we might expect to observe monotonicity among the \textit{ratios} of the parameters,
 $m_{match} / u_{match} \geq m_{partial-match} / u_{partial-match} \geq m_{non-match} / u_{non-match}$ when the comparison under consideration is ordinal (but not, in general, among the parameters themselves). In our application we do not impose this constraint although our parameter estimates, with $U$-correction, satisfy it approximately.

 \noindent {\bf Imposing the one-to-one constraint during or post-estimation.} 
 
 There are gains to enforcing the one-to-one constraint during estimation rather than post-hoc. For example, consider Table~\ref{tab:example_posterior_comparison}. The first three rows are ``William Brown''s found working as carpenters in 1932. The common names and occupation makes these individuals difficult to link. The remaining rows are records from 1936 found to have non-negligible matching probability to one of these. 
 
 The Bayesian model has no problem identifying the two exact matches to $a_2$ and $a_3$. Compare this to the fastLink estimated probabilities, which assigns significant matching probability between $a_2$ the first three 1936 candidates, despite the fact that there is an exact match present and the other two candidates differ on occupation, and one is missing an address and middle name. While fastLink's deduplication procedure makes the correct links here, the estimated matching probabilities are clearly nonsensical. And importantly, if the William Brown on Outlook Ave. had failed to register in 1936 then fastLink would have happily linked him to one of the other two records, as removing a single record pair has almost no influence on the estimated model parameters.  

\begin{table}
\centering
\resizebox{\textwidth}{!}{
\begin{tabular}{ccccccllllllrll}
\hline
\hline
&&&&& ID & Year & First & MI & Last & Female & Occupation & Street Number & Street Name & Street Type \\ 
\cline{6-15}
&&&&& $a_1$ & 1932 & william & e & brown &   0 & carpenter & 3200 & high & street \\ 
&&&&& $a_2$ & 1932 & william & e & brown &   0 & carpenter & 6401 & outlook & avenue \\ 
&&&&& $a_3$ & 1932 & william & j & broun &   0 & carpenter & 3026 & shattuck & avenue \\ 
\cline{6-15}
& &\\

\multicolumn{3}{c}{fastLink} & \multicolumn{3}{c}{Bayesian Model} \\
$a_1$ & $a_2$ & $a_3$ & $a_1$ & $a_2$ & $a_3$ \\
\cline{1-6}
1.00 & \textbf{1.00} & 0.00 & 0.00 & \textbf{1.00} & 0.00 & 1936 & william & e & brown &   0 & carpenter & 6401 & outlook & avenue \\ 
1.00 & 1.00 & 0.00 & 0.00 & 0.00 & 0.00 & 1936 & william & e & brown &   0 & title examiner & 6000 & romany & road \\ 
0.97 & 0.97 & 0.39 & 0.00 & 0.00 & 0.00 & 1936 & william & \textemdash & brown &   0 & musician & \textemdash & \textemdash & \textemdash \\ 
0.47 & 0.47 & 0.02 & 0.00 & 0.00 & 0.00 & 1936 & william & \textemdash & brown &   0 & laborer & \textemdash & decoto & \textemdash\\ 
0.21 & 0.21 & 0.01 & 0.00 & 0.00 & 0.00 & 1936 & william & h & brown &   0 & carpenter & \textemdash & san lean & \textemdash\\ 
0.03 & 0.05 & 0.00 & 0.00 & 0.00 & 0.00 & 1936 & william & \textemdash & brown &   0 & clerk & 25 & linda & avenue \\
0.03 & 0.05 & 0.00 & 0.00 & 0.00 & 0.00 & 1936 & william & \textemdash & brown &   0 & clerk & 2210 & tenth & avenue \\
0.01 & 0.01 & \textbf{1.00} & 0.00 & 0.00 & \textbf{1.00} & 1936 & william & j & brown &   0 & carpenter & 3026 & shattuck & avenue \\
0.01 & 0.01 & 0.00 & \textbf{0.64} & 0.00 & 0.00 & 1936 & william & m & brown &   0 & carpenter & 2205 & woolsey & street \\
0.00 & 0.00 & 0.99 & 0.00 & 0.00 & 0.00 & 1936 & william & j & brown &   0 & laundry businese & 1037 & oakland & avenue \\
0.00 & 0.00 & 0.99 & 0.00 & 0.00 & 0.00 & 1936 & william & j & brown &   0 & shoe salesman & 2200 & grant & street \\ 
0.00 & 0.00 & 0.99 & 0.00 & 0.00 & 0.00 & 1936 & william & j & brown &   0 & candy maker & 2311 & fifth & street \\ 
0.00 & 0.00 & 0.99 & 0.00 & 0.00 & 0.00 & 1936 & william & j & brown &   0 & laborer & 888 & fiftysecond & street \\ 
0.00 & 0.00 & 0.99 & 0.00 & 0.00 & 0.00 & 1936 & william & j & brown &   0 & clerk & 9223 & holly & street \\ 
\hline
\hline
\end{tabular}
}%
\caption{Example comparison of estimated posterior match probabilities from fastLink before deduplication and those estimated by the Bayesian model. Posterior match probabilities in bold indicate record pairs which are selected by the deduplication (fastLink) or contained in the Bayes estimator (Bayesian model).}
\label{tab:example_posterior_comparison}
\end{table}

\bibliographystyle{apa}
\bibliography{recordlinkage,assignment,misc,realignment}

\begin{thebibliography}{}

\bibitem[\protect\astroncite{Alicandro et~al.}{2017}]{alicandro2017differences}
Alicandro, G., Frova, L., Sebastiani, G., Boffetta, P., and La~Vecchia, C.
  (2017).
\newblock Differences in education and premature mortality: a record linkage
  study of over 35 million italians.
\newblock {\em European Journal of Public Health}.

\bibitem[\protect\astroncite{Andersen}{1979}]{andersen1979creation}
Andersen, K. (1979).
\newblock {\em The creation of a Democratic majority, 1928-1936}.
\newblock University of Chicago Press.

\bibitem[\protect\astroncite{Ball and Price}{2019}]{ball2019using}
Ball, P. and Price, M. (2019).
\newblock Using statistics to assess lethal violence in civil and inter-state
  war.
\newblock {\em Annual review of statistics and its application}, 6:63--84.

\bibitem[\protect\astroncite{Bertsekas}{1992}]{bertsekas1992auction}
Bertsekas, D.~P. (1992).
\newblock Auction algorithms for network flow problems: A tutorial
  introduction.
\newblock {\em Computational optimization and applications}, 1(1):7--66.

\bibitem[\protect\astroncite{Bertsekas}{1998}]{bertsekas1998network}
Bertsekas, D.~P. (1998).
\newblock {\em Network optimization: continuous and discrete models}.
\newblock Citeseer.

\bibitem[\protect\astroncite{Bertsekas and Eckstein}{1988}]{bertsekas1988dual}
Bertsekas, D.~P. and Eckstein, J. (1988).
\newblock Dual coordinate step methods for linear network flow problems.
\newblock {\em Mathematical Programming}, 42(1-3):203--243.

\bibitem[\protect\astroncite{Bertsekas and
  Tsitsiklis}{1989}]{bertsekas1989parallel}
Bertsekas, D.~P. and Tsitsiklis, J.~N. (1989).
\newblock {\em Parallel and distributed computation: numerical methods},
  volume~23.
\newblock Prentice hall Englewood Cliffs, NJ.

\bibitem[\protect\astroncite{Betancourt et~al.}{2016}]{betancourt2016}
Betancourt, B., Zanella, G., Miller, J.~W., Wallach, H., Zaidi, A., and
  Steorts, R.~C. (2016).
\newblock Flexible models for microclustering with application to entity
  resolution.
\newblock In Lee, D.~D., Sugiyama, M., Luxburg, U.~V., Guyon, I., and Garnett,
  R., editors, {\em Advances in Neural Information Processing Systems 29},
  pages 1417--1425. Curran Associates, Inc.

\bibitem[\protect\astroncite{Burkard et~al.}{2012}]{burkard2012assignment}
Burkard, R., Dell'Amico, M., and Martello, S. (2012).
\newblock {\em Assignment Problems, Revised Reprint}.
\newblock Other Titles in Applied Mathematics. Society for Industrial and
  Applied Mathematics.

\bibitem[\protect\astroncite{Carpaneto and Toth}{1983}]{carpaneto1983algorithm}
Carpaneto, G. and Toth, P. (1983).
\newblock Algorithm for the solution of the assignment problem for sparse
  matrices.
\newblock {\em Computing}, 31(1):83--94.

\bibitem[\protect\astroncite{Chen et~al.}{2018}]{chen2018}
Chen, B., Shrivastava, A., and Steorts, R.~C. (2018).
\newblock Unique entity estimation with application to the syrian conflict.
\newblock {\em Ann. Appl. Stat.}, 12(2):1039--1067.

\bibitem[\protect\astroncite{Christen}{2012}]{christen2012data}
Christen, P. (2012).
\newblock {\em Data matching: concepts and techniques for record linkage,
  entity resolution, and duplicate detection}.
\newblock Springer Science \& Business Media.

\bibitem[\protect\astroncite{Copas and Hilton}{1990}]{copas1990record}
Copas, J. and Hilton, F. (1990).
\newblock Record linkage: statistical models for matching computer records.
\newblock {\em Journal of the Royal Statistical Society. Series A (Statistics
  in Society)}, pages 287--320.

\bibitem[\protect\astroncite{Corder and Wolbrecht}{2016}]{corder2016counting}
Corder, J.~K. and Wolbrecht, C. (2016).
\newblock {\em Counting Women's Ballots: Female Voters from Suffrage through
  the New Deal}.
\newblock Cambridge University Press.

\bibitem[\protect\astroncite{Dalzell and Reiter}{2018}]{dalzell2018regression}
Dalzell, N.~M. and Reiter, J.~P. (2018).
\newblock Regression modeling and file matching using possibly erroneous
  matching variables.
\newblock {\em Journal of Computational and Graphical Statistics},
  27(4):728--738.

\bibitem[\protect\astroncite{Dalzell et~al.}{2017}]{dalzell2017file}
Dalzell, N.~M., Reiter, J.~P., and Boyd, G. (2017).
\newblock File matching with faulty continuous matching variables.
\newblock Working papers, U.S. Census Bureau, Center for Economic Studies.

\bibitem[\protect\astroncite{Dusetzina et~al.}{2014}]{dusetzina2014linking}
Dusetzina, S.~B., Tyree, S., Meyer, A.-M., Meyer, A., Green, L., and Carpenter,
  W.~R. (2014).
\newblock {\em Linking data for health services research: a framework and
  instructional guide}.
\newblock Agency for Healthcare Research and Quality (US), Rockville (MD).

\bibitem[\protect\astroncite{Enamorado et~al.}{2018a}]{fastLink_package}
Enamorado, T., Fifield, B., and Imai, K. (2018a).
\newblock {\em fastLink: Fast Probabilistic Record Linkage with Missing Data}.
\newblock R package version 0.5.0.

\bibitem[\protect\astroncite{Enamorado et~al.}{2018b}]{enamorado2018using}
Enamorado, T., Fifield, B., and Imai, K. (2018b).
\newblock Using a probabilistic model to assist merging of large-scale
  administrative records.
\newblock {\em American Political Science Review}, page 1–19.

\bibitem[\protect\astroncite{Erikson and Tedin}{1981}]{erikson1981partisan}
Erikson, R.~S. and Tedin, K.~L. (1981).
\newblock The 1928--1936 partisan realignment: The case for the conversion
  hypothesis.
\newblock {\em American Political Science Review}, 75(04):951--962.

\bibitem[\protect\astroncite{Fellegi and Sunter}{1969}]{fellegi1969theory}
Fellegi, I.~P. and Sunter, A.~B. (1969).
\newblock A theory for record linkage.
\newblock {\em Journal of the American Statistical Association},
  64(328):1183--1210.

\bibitem[\protect\astroncite{Fortini et~al.}{2001}]{fortini2001bayesian}
Fortini, M., Liseo, B., Nuccitelli, A., and Scanu, M. (2001).
\newblock On bayesian record linkage.
\newblock {\em Research in Official Statistics}, 4(1):185--198.

\bibitem[\protect\astroncite{Fortini et~al.}{2002}]{fortini2002modelling}
Fortini, M., Nuccitelli, A., Liseo, B., and Scanu, M. (2002).
\newblock Modelling issues in record linkage: a bayesian perspective.
\newblock In {\em Proceedings of the American Statistical Association, Survey
  Research Methods Section}, pages 1008--1013.

\bibitem[\protect\astroncite{Gazit}{1986}]{gazit1986optimal}
Gazit, H. (1986).
\newblock An optimal randomized parallel algorithm for finding connected
  components in a graph.
\newblock In {\em Foundations of Computer Science, 1986., 27th Annual Symposium
  on}, pages 492--501. IEEE.

\bibitem[\protect\astroncite{Green and Mardia}{2006}]{green2006bayesian}
Green, P.~J. and Mardia, K.~V. (2006).
\newblock Bayesian alignment using hierarchical models, with applications in
  protein bioinformatics.
\newblock {\em Biometrika}, 93(2):235--254.

\bibitem[\protect\astroncite{Gutman et~al.}{2013}]{gutman2013bayesian}
Gutman, R., Afendulis, C.~C., and Zaslavsky, A.~M. (2013).
\newblock A bayesian procedure for file linking to analyze end-of-life medical
  costs.
\newblock {\em Journal of the American Statistical Association},
  108(501):34--47.

\bibitem[\protect\astroncite{Hand and Christen}{2018}]{hand2018note}
Hand, D. and Christen, P. (2018).
\newblock A note on using the f-measure for evaluating record linkage
  algorithms.
\newblock {\em Statistics and Computing}, 28(3):539--547.

\bibitem[\protect\astroncite{Hof et~al.}{2017}]{hof2017probabilistic}
Hof, M.~H., Ravelli, A.~C., and Zwinderman, A.~H. (2017).
\newblock A probabilistic record linkage model for survival data.
\newblock {\em Journal of the American Statistical Association},
  112(520):1504--1515.

\bibitem[\protect\astroncite{Hong et~al.}{2016}]{hong2016solving}
Hong, C., Zhang, J., Chungfeng, C., and Qinyu, C. (2016).
\newblock Solving large-scale assignment problems by kuhn-munkres algorithm.
\newblock In {\em 2nd Int. Conf. Adv. Mech. Eng. Ind. Informatics (AMEII 2016),
  no. Ameii}, pages 822--827.

\bibitem[\protect\astroncite{Jaro}{1989}]{jaro1989advances}
Jaro, M.~A. (1989).
\newblock Advances in record-linkage methodology as applied to matching the
  1985 census of tampa, florida.
\newblock {\em Journal of the American Statistical Association},
  84(406):414--420.

\bibitem[\protect\astroncite{Jonker and Volgenant}{1987}]{jonker1987shortest}
Jonker, R. and Volgenant, A. (1987).
\newblock A shortest augmenting path algorithm for dense and sparse linear
  assignment problems.
\newblock {\em Computing}, 38(4):325--340.

\bibitem[\protect\astroncite{Jonker and Volgenant}{1986}]{jonker1986improving}
Jonker, R. and Volgenant, T. (1986).
\newblock Improving the hungarian assignment algorithm.
\newblock {\em Operations Research Letters}, 5(4):171--175.

\bibitem[\protect\astroncite{Kuhn}{1955}]{kuhn1955hungarian}
Kuhn, H.~W. (1955).
\newblock The hungarian method for the assignment problem.
\newblock {\em Naval Research Logistics (NRL)}, 2(1-2):83--97.

\bibitem[\protect\astroncite{Ladd and Hadley}{1975}]{ladd1975transformations}
Ladd, E.~C. and Hadley, C.~D. (1975).
\newblock {\em Transformations of the American party system: Political
  coalitions from the New Deal to the 1970s}.
\newblock WW Norton.

\bibitem[\protect\astroncite{Larsen}{2005}]{larsen2005advances}
Larsen, M.~D. (2005).
\newblock Advances in record linkage theory: Hierarchical bayesian record
  linkage theory.
\newblock In {\em Proceedings of the American Statistical Association, Survey
  Research Methods Section}, pages 3277--3284.

\bibitem[\protect\astroncite{Larsen}{2010}]{larsen2010record}
Larsen, M.~D. (2010).
\newblock Record linkage modeling in federal statistical databases.
\newblock In {\em FCSM Research Conference, Washington, DC. Federal Committee
  on Statistical Methodology}. Citeseer.

\bibitem[\protect\astroncite{Larsen and Rubin}{2001}]{larsen2001iterative}
Larsen, M.~D. and Rubin, D.~B. (2001).
\newblock Iterative automated record linkage using mixture models.
\newblock {\em Journal of the American Statistical Association},
  96(453):32--41.

\bibitem[\protect\astroncite{Lawler}{1976}]{lawler1976combinatorial}
Lawler, E.~L. (1976).
\newblock {\em Combinatorial optimization: networks and matroids}.
\newblock Courier Corporation.

\bibitem[\protect\astroncite{Liseo and Tancredi}{2011}]{liseo2011bayesian}
Liseo, B. and Tancredi, A. (2011).
\newblock Bayesian estimation of population size via linkage of multivariate
  normal data sets.
\newblock {\em Journal of Official Statistics}, 27(3):491--505.

\bibitem[\protect\astroncite{Mackay et~al.}{2015}]{mackay2015educational}
Mackay, D.~F., Wood, R., King, A., Clark, D.~N., Cooper, S.-A., Smith, G.~C.,
  and Pell, J.~P. (2015).
\newblock Educational outcomes following breech delivery: a record-linkage
  study of 456 947 children.
\newblock {\em International journal of epidemiology}, 44(1):209--217.

\bibitem[\protect\astroncite{Marchant et~al.}{2019}]{marchant2019d}
Marchant, N.~G., Steorts, R.~C., Kaplan, A., Rubinstein, B.~I., and Elazar,
  D.~N. (2019).
\newblock d-blink: Distributed end-to-end bayesian entity resolution.
\newblock {\em arXiv preprint arXiv:1909.06039}.

\bibitem[\protect\astroncite{McVeigh and Murray}{2017}]{mcveigh2017practical}
McVeigh, B.~S. and Murray, J.~S. (2017).
\newblock Practical bayesian inference for record linkage.

\bibitem[\protect\astroncite{Murray}{2016}]{murray2016probabilistic}
Murray, J.~S. (2016).
\newblock Probabilistic record linkage and deduplication after indexing,
  blocking, and filtering.
\newblock {\em Journal of Privacy and Confidentiality}, 7.

\bibitem[\protect\astroncite{Newcombe et~al.}{1959}]{newcombe1959automatic}
Newcombe, H.~B., Kennedy, J.~M., Axford, S., and James, A.~P. (1959).
\newblock Automatic linkage of vital records.
\newblock {\em Science}, 130(3381):954--959.

\bibitem[\protect\astroncite{Norpoth}{2019}]{norpoth2019american}
Norpoth, H. (2019).
\newblock The american voter in 1932: Evidence from a confidential survey.
\newblock {\em PS: Political Science \& Politics}, 52(1):14--19.

\bibitem[\protect\astroncite{Orlin and Lee}{1993}]{orlin1993quickmatch}
Orlin, J.~B. and Lee, Y. (1993).
\newblock Quickmatch--a very fast algorithm for the assignment problem.
\newblock Working Paper 3547-93, Sloan School of Management, Massachusetts
  Institute of Technology, Cambridge, MA.

\bibitem[\protect\astroncite{Price et~al.}{2014}]{price2014updated}
Price, M., Gohdes, A., and Ball, P. (2014).
\newblock Updated statistical analysis of documentation of killings in the
  syrian arab republic.
\newblock {\em Human Rights Data Analysis Group, Geneva}.

\bibitem[\protect\astroncite{Rice}{2006}]{rice2006mathematical}
Rice, J.~A. (2006).
\newblock {\em Mathematical Statistics and Data Analysis}, chapter~7.
\newblock Belmont, CA: Duxbury Press, third edition.

\bibitem[\protect\astroncite{Sadinle}{2013}]{sadinle2013bayesian}
Sadinle, M. (2013).
\newblock A bayesian framework for duplicate detection, record linkage, and
  subsequent inference with linked files.
\newblock {\em Under review}.

\bibitem[\protect\astroncite{Sadinle}{2014}]{sadinle2014detecting}
Sadinle, M. (2014).
\newblock Detecting duplicates in a homicide registry using a bayesian
  partitioning approach.
\newblock {\em The Annals of Applied Statistics}, 8(4):2404--2434.

\bibitem[\protect\astroncite{Sadinle}{2017}]{sadinle2017bayesian}
Sadinle, M. (2017).
\newblock Bayesian estimation of bipartite matchings for record linkage.
\newblock {\em Journal of the American Statistical Association},
  112(518):600--612.

\bibitem[\protect\astroncite{Sadinle}{2018}]{sadinle2018bayesian}
Sadinle, M. (2018).
\newblock Bayesian propagation of record linkage uncertainty into population
  size estimation of human rights violations.
\newblock {\em The Annals of Applied Statistics}, 12(2):1013--1038.

\bibitem[\protect\astroncite{Sauleau et~al.}{2005}]{sauleau2005medical}
Sauleau, E.~A., Paumier, J.-P., and Buemi, A. (2005).
\newblock Medical record linkage in health information systems by approximate
  string matching and clustering.
\newblock {\em BMC Medical Informatics and Decision Making}, 5(1):32.

\bibitem[\protect\astroncite{Spahn}{2017}]{spahn2018before}
Spahn, B. (2017).
\newblock Before the american voter.
\newblock {\em Available at SSRN: https://www.ssrn.com/abstract=3478473}.

\bibitem[\protect\astroncite{Steorts et~al.}{2015}]{steorts2015entity}
Steorts, R.~C. et~al. (2015).
\newblock Entity resolution with empirically motivated priors.
\newblock {\em Bayesian Analysis}, 10(4):849--875.

\bibitem[\protect\astroncite{Steorts et~al.}{2016}]{steorts2016bayesian}
Steorts, R.~C., Hall, R., and Fienberg, S.~E. (2016).
\newblock A bayesian approach to graphical record linkage and deduplication.
\newblock {\em Journal of the American Statistical Association},
  111(516):1660--1672.

\bibitem[\protect\astroncite{Steorts et~al.}{2014}]{steorts2014comparison}
Steorts, R.~C., Ventura, S.~L., Sadinle, M., and Fienberg, S.~E. (2014).
\newblock A comparison of blocking methods for record linkage.
\newblock In {\em International Conference on Privacy in Statistical
  Databases}, pages 253--268. Springer.

\bibitem[\protect\astroncite{Sundquist}{1983}]{sundquist1983dynamics}
Sundquist, J.~L. (1983).
\newblock {\em Dynamics of the party system}.
\newblock Brookings Institute Washington, DC.

\bibitem[\protect\astroncite{Tancredi et~al.}{2013}]{tancredi2013accounting}
Tancredi, A., Auger-M{\'e}th{\'e}, M., Marcoux, M., and Liseo, B. (2013).
\newblock Accounting for matching uncertainty in two stage capture--recapture
  experiments using photographic measurements of natural marks.
\newblock {\em Environmental and ecological statistics}, 20(4):647--665.

\bibitem[\protect\astroncite{Tancredi et~al.}{2011}]{tancredi2011hierarchical}
Tancredi, A., Liseo, B., et~al. (2011).
\newblock A hierarchical bayesian approach to record linkage and population
  size problems.
\newblock {\em The Annals of Applied Statistics}, 5(2B):1553--1585.

\bibitem[\protect\astroncite{Tancredi et~al.}{2018}]{tancredi2018}
Tancredi, A., Steorts, R., and Liseo, B. (2018).
\newblock A unified framework for de-duplication and population size
  estimation.
\newblock {\em Bayesian Anal.}
\newblock Advance publication.

\bibitem[\protect\astroncite{Tarjan}{1972}]{tarjan1972depth}
Tarjan, R. (1972).
\newblock Depth-first search and linear graph algorithms.
\newblock {\em SIAM journal on computing}, 1(2):146--160.

\bibitem[\protect\astroncite{Thibaudeau}{1993}]{thibaudeau1993discrimination}
Thibaudeau, Y. (1993).
\newblock The discrimination power of dependency structures in record linkage.
\newblock {\em Survey Methodology}, 19:31--38.

\bibitem[\protect\astroncite{Winkler et~al.}{2010}]{winkler2010fast}
Winkler, W., Yancey, W., and Porter, E. (2010).
\newblock Fast record linkage of very large files in support of decennial and
  administrative records projects.
\newblock In {\em Proceedings of the Section on Survey Research Methods,
  American Statistical Association}, pages 2120--2130.

\bibitem[\protect\astroncite{Winkler}{1988}]{winkler1988using}
Winkler, W.~E. (1988).
\newblock Using the em algorithm for weight computation in the fellegi-sunter
  model of record linkage.
\newblock In {\em Proceedings of the Section on Survey Research Methods,
  American Statistical Association}, pages 667--671.

\bibitem[\protect\astroncite{Winkler}{1990}]{winkler1990string}
Winkler, W.~E. (1990).
\newblock String comparator metrics and enhanced decision rules in the
  fellegi-sunter model of record linkage.
\newblock In {\em Proceedings of the Section on Survey Research Methods,
  American Statistical Association}, pages 354--359.

\bibitem[\protect\astroncite{Winkler}{1993}]{winkler1993improved}
Winkler, W.~E. (1993).
\newblock Improved decision rules in the fellegi-sunter model of record
  linkage.
\newblock In {\em Proceedings of the Section on Survey Research Methods,
  American Statistical Association}, pages 274--279.

\bibitem[\protect\astroncite{Winkler}{2002}]{winkler2002methods}
Winkler, W.~E. (2002).
\newblock Methods for record linkage and bayesian networks.
\newblock Technical report, Statistical Research Division, US Census Bureau,
  Washington, DC.

\bibitem[\protect\astroncite{Winkler and
  Thibaudeau}{1991}]{winkler1991application}
Winkler, W.~E. and Thibaudeau, Y. (1991).
\newblock An application of the fellegi-sunter model of record linkage to the
  1990 us decennial census.
\newblock {\em US Bureau of the Census}, pages 1--22.

\bibitem[\protect\astroncite{Yancey}{2002}]{yancey2002bigmatch}
Yancey, W.~E. (2002).
\newblock Bigmatch: A program for extracting probable matches from a large file
  for record linkage.
\newblock Technical report statistical research report series rrc2002/01, U.S.
  Bureau of the Census, Washington, D.C.

\bibitem[\protect\astroncite{Zanella}{2019}]{zanella2019informed}
Zanella, G. (2019).
\newblock Informed proposals for local mcmc in discrete spaces.
\newblock {\em Journal of the American Statistical Association}, 0(0):1--27.

\end{thebibliography}


\begin{thebibliography}{}

\bibitem[\protect\astroncite{Bertsekas}{1992}]{bertsekas1992auction}
Bertsekas, D.~P. (1992).
\newblock Auction algorithms for network flow problems: A tutorial
  introduction.
\newblock {\em Computational optimization and applications}, 1(1):7--66.

\bibitem[\protect\astroncite{Bertsekas}{1998}]{bertsekas1998network}
Bertsekas, D.~P. (1998).
\newblock {\em Network optimization: continuous and discrete models}.
\newblock Citeseer.

\bibitem[\protect\astroncite{Bertsekas and Eckstein}{1988}]{bertsekas1988dual}
Bertsekas, D.~P. and Eckstein, J. (1988).
\newblock Dual coordinate step methods for linear network flow problems.
\newblock {\em Mathematical Programming}, 42(1-3):203--243.

\bibitem[\protect\astroncite{Bertsekas and
  Tsitsiklis}{1989}]{bertsekas1989parallel}
Bertsekas, D.~P. and Tsitsiklis, J.~N. (1989).
\newblock {\em Parallel and distributed computation: numerical methods},
  volume~23.
\newblock Prentice hall Englewood Cliffs, NJ.

\bibitem[\protect\astroncite{Carpaneto and Toth}{1983}]{carpaneto1983algorithm}
Carpaneto, G. and Toth, P. (1983).
\newblock Algorithm for the solution of the assignment problem for sparse
  matrices.
\newblock {\em Computing}, 31(1):83--94.

\bibitem[\protect\astroncite{Enamorado et~al.}{2018}]{enamorado2018using}
Enamorado, T., Fifield, B., and Imai, K. (2018).
\newblock Using a probabilistic model to assist merging of large-scale
  administrative records.
\newblock {\em American Political Science Review}, page 1–19.

\bibitem[\protect\astroncite{Fortini et~al.}{2001}]{fortini2001bayesian}
Fortini, M., Liseo, B., Nuccitelli, A., and Scanu, M. (2001).
\newblock On bayesian record linkage.
\newblock {\em Research in Official Statistics}, 4(1):185--198.

\bibitem[\protect\astroncite{Fortini et~al.}{2002}]{fortini2002modelling}
Fortini, M., Nuccitelli, A., Liseo, B., and Scanu, M. (2002).
\newblock Modelling issues in record linkage: a bayesian perspective.
\newblock In {\em Proceedings of the American Statistical Association, Survey
  Research Methods Section}, pages 1008--1013.

\bibitem[\protect\astroncite{Green and Mardia}{2006}]{green2006bayesian}
Green, P.~J. and Mardia, K.~V. (2006).
\newblock Bayesian alignment using hierarchical models, with applications in
  protein bioinformatics.
\newblock {\em Biometrika}, 93(2):235--254.

\bibitem[\protect\astroncite{Hong et~al.}{2016}]{hong2016solving}
Hong, C., Zhang, J., Chungfeng, C., and Qinyu, C. (2016).
\newblock Solving large-scale assignment problems by kuhn-munkres algorithm.
\newblock In {\em 2nd Int. Conf. Adv. Mech. Eng. Ind. Informatics (AMEII 2016),
  no. Ameii}, pages 822--827.

\bibitem[\protect\astroncite{Jaro}{1989}]{jaro1989advances}
Jaro, M.~A. (1989).
\newblock Advances in record-linkage methodology as applied to matching the
  1985 census of tampa, florida.
\newblock {\em Journal of the American Statistical Association},
  84(406):414--420.

\bibitem[\protect\astroncite{Jonker and Volgenant}{1987}]{jonker1987shortest}
Jonker, R. and Volgenant, A. (1987).
\newblock A shortest augmenting path algorithm for dense and sparse linear
  assignment problems.
\newblock {\em Computing}, 38(4):325--340.

\bibitem[\protect\astroncite{Jonker and Volgenant}{1986}]{jonker1986improving}
Jonker, R. and Volgenant, T. (1986).
\newblock Improving the hungarian assignment algorithm.
\newblock {\em Operations Research Letters}, 5(4):171--175.

\bibitem[\protect\astroncite{Kuhn}{1955}]{kuhn1955hungarian}
Kuhn, H.~W. (1955).
\newblock The hungarian method for the assignment problem.
\newblock {\em Naval Research Logistics (NRL)}, 2(1-2):83--97.

\bibitem[\protect\astroncite{Larsen}{2005}]{larsen2005advances}
Larsen, M.~D. (2005).
\newblock Advances in record linkage theory: Hierarchical bayesian record
  linkage theory.
\newblock In {\em Proceedings of the American Statistical Association, Survey
  Research Methods Section}, pages 3277--3284.

\bibitem[\protect\astroncite{Larsen}{2010}]{larsen2010record}
Larsen, M.~D. (2010).
\newblock Record linkage modeling in federal statistical databases.
\newblock In {\em FCSM Research Conference, Washington, DC. Federal Committee
  on Statistical Methodology}. Citeseer.

\bibitem[\protect\astroncite{Lawler}{1976}]{lawler1976combinatorial}
Lawler, E.~L. (1976).
\newblock {\em Combinatorial optimization: networks and matroids}.
\newblock Courier Corporation.

\bibitem[\protect\astroncite{Orlin and Lee}{1993}]{orlin1993quickmatch}
Orlin, J.~B. and Lee, Y. (1993).
\newblock Quickmatch--a very fast algorithm for the assignment problem.
\newblock Working Paper 3547-93, Sloan School of Management, Massachusetts
  Institute of Technology, Cambridge, MA.

\bibitem[\protect\astroncite{Sadinle}{2017}]{sadinle2017bayesian}
Sadinle, M. (2017).
\newblock Bayesian estimation of bipartite matchings for record linkage.
\newblock {\em Journal of the American Statistical Association},
  112(518):600--612.

\bibitem[\protect\astroncite{Tancredi et~al.}{2011}]{tancredi2011hierarchical}
Tancredi, A., Liseo, B., et~al. (2011).
\newblock A hierarchical bayesian approach to record linkage and population
  size problems.
\newblock {\em The Annals of Applied Statistics}, 5(2B):1553--1585.

\bibitem[\protect\astroncite{Tarjan}{1972}]{tarjan1972depth}
Tarjan, R. (1972).
\newblock Depth-first search and linear graph algorithms.
\newblock {\em SIAM journal on computing}, 1(2):146--160.

\end{thebibliography}
\end{document}